**Brain State Control by Closed-Loop Environmental Feedback**

Christopher L. Buckley[1,2,*], Satohiro Tajima[1], Toru Yanagawa[1], Kana Takakura[1], Yasuo Nagasaka[1], Naotaka Fujii[1], and Taro Toyoizumi[1,*]

[1] RIKEN Brain Science Institute, Japan

[2] University of Sussex, UK

**Summary**

Brain state regulates sensory processing and motor control for adaptive behavior. Internal mechanisms of brain state control are well studied, but the role of external modulation from the environment is not well understood. Here, we examined the role of closed-loop environmental (CLE) feedback, in comparison to open-loop sensory input, on brain state and behavior in diverse vertebrate systems. In fictively swimming zebrafish, CLE feedback for optomotor stability controlled brain state by reducing coherent neuronal activity. The role of CLE feedback in brain state was also shown in a model of rodent active whisking, where brief interruptions in this feedback enhanced signal-to-noise ratio for detecting touch. Finally, in monkey visual fixation, artificial CLE feedback suppressed stimulus-specific neuronal activity and improved behavioral performance. Our findings show that the environment mediates continuous closed-loop feedback that controls neuronal gain, regulating brain state, and that brain function is an emergent property of brain-environment interactions.

**Introduction**

The repertoire of animal behavior involves both passive and active interactions of the brain with the environment. Passive interactions are driven by the environment and convey signals for passive sensing and alert. Active interactions, in contrast, are bidirectional, mediating movement and goal-directed behaviors. During active interactions sensory input from the environment is shaped by motor actions, and sensory percepts inform future motor commands, forming a closed-loop between action and perception. This closed-loop environmental (CLE) feedback is central to the production of motor control behaviors (Wolpert and Ghahramani, 2000; Kawato, 1999; Scott, 2004) and active sensing behaviors such as saccading and sniffing (Smith 1962; Von Holst, 1954; Crapse and Sommer, 2008). Yet, how this CLE feedback impacts on brain state, i.e., brain-wide neuronal dynamics and processing, has received relatively little attention.

Brain state, typically characterized by the degree of synchronously fluctuating neuronal activity as reported by electroencephalography (EEG), local field potentials (LFP), electrocortiography (ECoG), membrane potential, and population spiking activity, is strongly modulated by behavioral context (Buzsáki et al., 2012; Harris and Thiele, 2011). One of the earliest experimental examples of this phenomenon was the demonstration that alpha frequency power (7.5–12.5Hz) in EEG recordings is enhanced when subjects close their eyes or during periods of drowsiness (Berger, 1929). More generally recent experiments in rodents have demonstrated that synchronous low-frequency fluctuations (a *synchronized state*) are typically suppressed (moving to a *desynchronized state*) when animals transition from quiet attentive behaviors, where animals are largely passive, to active behaviors such as running or whisking (Crochet and Petersen, 2006; Niel and Stryker, 2010; Schneider et al., 2014; Otazu et al., 2009; Zagha et al., 2013).

Brain states not only alter spontaneous brain dynamics, but also sensory representations, presumably allowing brain function to adapt to behavioral context. For example, it has been shown that the onset of running amplifies cortical responses to visual and auditory stimuli (Niell and Stryker, 2010, Fu et al., 2014; Schneider et al., 2014; McGinley and McCormick, 2014). In addition, cortical responses to punctuate stimuli are larger during synchronized states (Poulet and Petersen, 2008; Fanselow and Nicolelis, 1999; Castro-Alamancos, 2004, Otazu et al., 2009). These brain state changes are typically thought to happen across the whole brain (Berger, 1929, Steriade, 2001; Fu et al., 2014; McGinley and McCormick, 2014) and are cross-modal (Otazu et al., 2009).

Several internal mechanisms have been implicated in these changes including neuromodulatory pathways (Goard and Dan, 2009; Polack et al., 2012; Pinto et al., 2013; Fu et al., 2014), thalamo-cortical projections (Poulet et al., 2012), and intracortical feedback (Zagha et al., 2014; Schneider et al., 2014). Changes in the external sensory environment also play an important role. Under passive sensing conditions sensory stimulation can shift neural activity in the visual cortex between asynchronous and synchronous regimes (Tan et al., 2014) and more generally sensory stimuli quench neural variability (Churchland et al., 2010). More relevant to the present study, brain state transitions often coincide with the onset of active behaviors, which are characterized by the presence of *reafferent* input (sensory input resulting from one's own action, mediated by the environment). It is well known that reafferent input strongly influences neural activity (Curtis and Kleinfeld, 2009; Urbain et al., 2015; Brooks et al., 2015) and modulate sensory responses (Diamond et al., 2008). However, the role played by reafferent input in regulating brain state is not well understood. In this paper, we combine theoretical and experimental approaches to dissect the effect CLE feedback mediated by reafferent input, as oppose to sensory input per se, has on modulating brain state. In doing so we elucidate a novel function that CLE feedback plays in sensing and behavior.

**Results**

During active behavior, sensory input to the brain is directly shaped by motor actions, and reciprocally, motor actions are informed by prior sensations forming a closed-loop between the brain and the environment. Under such conditions, it is possible to distinguish two sources of sensory input (Von Holst, 1954), see Figure 1A. First, *exafferent* input describes sensory signals that originate in the environment, but which are completely external to the brain (Von Holst, 1954). Second, *reafferent* input describes sensory signals, which, while mediated by the environment, are a direct consequence of an animal's own actions and constitute a *sensory feedback* signal to the brain (Von Holst, 1954). While it is known that sensory input in general can influence brain state variability (Churchland et al., 2010), it is not clear whether exafferent and reafferent input play distinct roles in determining these changes, or whether only total synaptic input is of importance. To examine the effect of each type of input on brain state, we distinguish three types of brain-environment interaction. First, an *open-loop* condition, where sensory input is not coupled to motor output and thus the brain receives sensory input independent of its own activity (Figure 1B). In this condition, reafferent input is absent and the brain receives only exafferent input. Second, a *closed-loop* condition, where the brain interacts with the environment in a closed-loop thus coupling motor action and sensory perception (Figure 1C). In this condition, the brain receives reafferent input in addition to any possible exafferent input. Lastly, we define a *replay condition,* where again the brain operates in an open-loop but where reafferent input experienced during a prior closed-loop condition is recorded and replayed as exafferent input (Figure 1D). In this condition, the total sensory input is identical to that in the closed-loop condition, but reafferent input is absent. Specifically, the sensory input during the replay condition is not a consequence of motor actions within that condition. Therefore, this condition serves as a strong control allowing us to distinguish precisely the contribution made by reafferent input rather than that of total sensory input, to the brain state.

To understand the possible effects that CLE feedback could have on brain activity we introduce a simple idealized model. We start by describing collective neuronal activity, e.g., an EEG signal, of an open-loop brain (denoted as $B_o(t)$), in terms of a first-order differential equation,

$$\frac{dB_o(t)}{dt} = -\frac{B_o(t)}{\tau} + \xi_o(t),$$

where $\xi_o$ is white noise of instantaneous variance $\sigma^2$ generated inside the brain, $t$ is time, and $\tau$ is the time constant of the system. The autocorrelation peak (instantaneous variance) of variable $B_o$ is given by $\text{Peak}_o = \sigma^2\tau/2$ (see, Figure 1B for open-loop time trace). To describe neuronal activity in the closed-loop condition (denoted as $B_c(t)$), we approximate reafferent input as a linear function of the brain variable, i.e.,

$$\frac{dB_c(t)}{dt} = -\frac{B_c(t)}{\tau} + wB_c(t) + \xi_c(t),$$

where $w$ scales the strength of reafferent input and $\xi_c$ is again white noise of instantaneous variance $\sigma^2$. (Later we also explore a more realistic reafferent input that involves filtering and delay). In this condition, the continuous cycles of reafferent input constitutes a CLE feedback signal to the brain. The presence of this CLE feedback changes the effective time constant to $\tau_{\text{eff}} = \tau/(1 - w\tau)$ or, equivalently, it the changes the system's gain. We can quantify this change by using an expression for the autocorrelation peak of the system, as follows, $\text{Peak}_c = \text{Peak}_o/(1 - w\tau)$. In particular, we find that CLE feedback suppresses fluctuations within the brain if this feedback is negative ($w < 0$; see Figure 1C for closed-loop time trace). In contrast, we can describe neuronal activity in the replay condition (denoted as $B_r(t)$) as:

$$\frac{dB_r(t)}{dt} = -\frac{B_r(t)}{\tau} + wB_c(t) + \xi_r(t),$$

where $\xi_r$ is again white noise of instantaneous variance $\sigma^2$. Here, the brain receives the same total sensory input as in the closed-loop condition, $wB_c(t)$, but as exafferent input, i.e., it depends on the dynamics of $B_c(t)$, and not on $B_r(t)$. Again we can calculate the autocorrelation peak of the brain variable in this condition as $\text{Peak}_r = \text{Peak}_c + \text{Peak}_o * 2w\tau/(w\tau - 2)$. Notably, even though the brain receives exactly the same total sensory input in the replay and closed-loop conditions the amplitudes of fluctuations are not the same. In particular, if CLE feedback is negative, we obtain $\text{peak}_c < \text{peak}_o < \text{peak}_r$ (see, Figure 1D for replay time trace). In summary, this simple model suggests that CLE feedback constituted by continuous reafferent input could have a profound effect on neuronal fluctuations and thus could be implicated in the modulation of brain state. In the next section, we examine this hypothesis using experimental data.

Negative CLE feedback suppresses synchronous neural fluctuations

We tested our predictions by analysing two-photon imaging data recorded from larval zebrafish behaving in a virtual flow simulator (Ahrens et al., 2012), a setup that allowed us to quantify the differences between the closed-loop and replay conditions directly (Figure 1). In this setup, bouts of activity recorded from motor nerves along the spine of a paralyzed fish were translated into a backward drift of a visual grating, simulating forward swimming. This allows the fish to maintain its perceived horizontal position by swimming against water flow (Ahrens et al., 2012), (Figure 2A). In a transgenic fish expressing the calcium indicator GCaMP2 brain-wide calcium activity was monitored using a two-photon microscope to scan single planes in the brain. In a closed-loop condition, the fish actively maintain their position in the virtual environment. In a replay condition, the same fish receives a replay of the closed-loop visual stimulus without real-time visual feedback (Figure 2A), (Ahrens et al., 2012). We found that neural dynamics between the two conditions were significantly different despite the fact that both the closed-loop and replay conditions involve identical sensory input (Figure 2B). Notably, the only information the fish received about oncoming flow was visual, i.e., there was no proprioceptive input as the fish was paralyzed (Ahrens et al., 2012). Individual neurons were heterogeneous across the whole brain, but on average low frequency (0.01 - 0.15 Hz) fluctuations were suppressed ($p = 0.046$, sign test) and neurons were decorrelated ($p = 0.005$, sign test) under the closed-loop condition compared to the replay condition (Figure 2B). Changes in the geometric mean of low frequency fluctuations and correlation were highly correlated in each pair of cells ($r=0.69$, $p<10^{-10}$, Spearman's rank correlation), suggesting a common cause (see Figure 2C). The decorrelation effect was not an artifact of measurement noise, which may dominate at high frequency because the result was robust to the removal of low-level calcium activity by thresholding (see Supplementary Figure S1A). This change aligns with brain state transitions at the onset of active behavior that has been reported in other species (see Poulet et al., 2012 or Harris and Thiele, 2011 for a review). This suppression of low frequency power and correlation could be caused by an efference copy signal (signal encoding intended motor action) that suppresses synchronous neuronal fluctuations (Zagha et al., 2014; Schneider et al., 2014). However, while motor activity levels were higher in the closed-loop condition than the replay condition, this did not explain the difference in neuronal fluctuations. Specifically, the motor activity level positively correlated with the changes in low frequency power (Supplementary Figure S1B, top panel; $r = 0.18$, $p < 10^{-2}$, Spearman's rank correlation) and was not significantly correlated with changes in pairwise correlations between cells (Supplementary Figure S1B, bottom panel $r = 0.03$, $p > 0.5$, Spearman's rank correlation).

The only difference between the closed-loop and replay conditions was the presence of CLE feedback; thus it can be assumed that this must play a causal role in these changes. To investigate this further, we quantified CLE feedback by first estimating how fish behavior affects neural activity, and conversely, how neural activity drives their behavior. To this end we calculated linear filters that characterize dynamic interactions between single neuron activity (the brain variable: $B$) and swimming power as quantified by the activity of motor neurons (a putative environmental variable: $E$) (see Methods). Figure 3A (inset) schematically shows their interaction under each condition. In the closed-loop condition $B$ and $E$ interact mutually; this has the potential to confound calculation of independent causal filters. Consequently, we calculated filters based on the replay condition (however, see below for filters computed based on the closed-loop condition). In this condition, the brain variable $B'$ is driven by the recorded CLE variable $E$, which in turn generates its own environmental variable $E'$, without involving CLE feedback. Specifically, for each observed neuron, we computed a linear filter that predicts $B'$ from $E$ and a linear filter that predicts $E'$ from $B'$. While these filters are neuron-dependent, the average normalized filters showed clear net effects (Figure 3A). On average, across cells, the interaction from the brain to the environment, $B' \rightarrow E'$ (Figure 3A, purple solid line), was strongly positive, indicating that an increase in neural activity positively drove motor behavior. However, we found that interaction from the environment to the brain, *E→B'* (Figure 3A, pink solid line), was net negative, indicating that an increase in motor behavior on average suppressed neural activity. It is reasonable to assume that the same filters (circuits) operate in the closed-loop condition. If this is the case, CLE feedback for each cell in the closed-loop condition is characterized by the convolution of the two filters, i.e., by the convolution of the *B'→E'* and *E→B'* interactions computed for each cell. The convolution of these filters was on average also negative (Figure 3A, cyan dashed line), peaking at about 1 s. This suggests, that on average, increases in neural activity self suppressed after 1 s due to negative CLE feedback. Notably, the negative CLE feedback interactions were also confirmed by behavioral analysis, in which the the *E→E'* filter was estimated directly (see Supplementary Figure S1C), which supported the robustness of our results.

Next, we considered what the consequence of this negative CLE feedback on each cell's activity would be. Our conceptual model, Figure 1, suggests that fluctuations in neural activity should decrease if a cell receives negative CLE feedback. In this case the estimated CLE feedback in each cell should predict to what extent fluctuations in low frequency power should change in the closed-loop condition relative to the replay condition (see Methods). As theoretically expected, Figure 3B demonstrates that the predicted degree to which a neuron's activity was suppressed during the closed-loop condition relative to the replay condition was highly correlated with what was actually observed (r = 0.39, $p < 10^{-8}$, Spearman's rank correlation). Note that neural activity in the closed-loop condition was not used to fit each filter. Although the majority of cells across the fish brain were suppressed by the behavior, the top 10 percentile of cells that were both strongly suppressed, and strongly involved in, negative CLE feedback were clustered in the cerebellum (Figure 3C), a brain area implicated in sensory-motor planning and coordination (Scott, 2004). This supports the idea that the cerebellum plays a central role mediating negative closed-loop interaction between the brain and the environment by converting sensation into action in fish during optic flow stabilization.

We then investigated whether the dynamic relation between neuronal activity and motor activity were also affected by CLE feedback. To do this, we quantified the brain/environment dynamics for each neuron by naively evaluating the *E→B* filter in the closed-loop condition, and compared this to the corresponding *E→B'* filter in the replay condition (Figure 3D). These two filters were generally distinct in the observed neurons, but were particularly so for those cells that were strongly stabilized by CLE feedback (Figure 3D; Inset). In these neurons, while the *E→B'* filter from the replay condition was approximately causal, as expected, the closed-loop filter *E→B* had an additional acausal component. We suggest that this latter filter reflects the closed-loop interaction between the brain *B* and the environment *E*. To test if the closed-loop interaction could explain this discrepancy, we theoretically predicted the *E→B* filter in the closed-loop condition based on data from the replay condition, i.e., using both the *E→B'* and *B'→E'* filters (see Methods). We found that, on average, this closed-loop effect could account for the difference between *E→B* and *E→B'* as calculated for the closed-loop stabilized cells (Figure 3D; Inset). To quantify this for each cell, we calculated the mean square error between the predicted and *E→B* filters and between the *E→B* and *E→B'* filter. The ratio of these two error terms then quantifies the fraction of the mean square error that was explained by the prediction. This ratio was significantly less than one (median = 0.80, $p < 10^{-11}$, sign test), indicating that the prediction was more accurate when the closed-loop effect was included, and the ratio was positively correlated (r=0.25, $p<10^{-13}$, Spearman's rank correlation) with the degree to which individual cells were stabilized in the closed-loop condition (Figure 3D). Altogether, these results indicated that neuronal dynamics, as well as its relation to sensory stimulus and behavior, not only depend on brain circuits, but are also dynamically shaped by the mutual interaction between the brain and the environment.

**CLE feedback explains the difference between active and passive sensing**

Behaviors not only induce brain state transitions, but also differentiate active sensing from passive sensing. To investigate whether these differences could be accounted for by CLE feedback, we examined a well-studied model of a brain state transition in the rodent whisker system. Specifically, we consider CLE feedback to the rodent brain mediated by whisking vibrissa. Although more precisely we should refer to this as closed-loop body/environmental feedback, we retain CLE acronym by generalizing the notion of environment to refer to all processes outside of the brain highlighting the generality of our theory. Whisking not only changes brain state (Crochet and Petersen, 2006; Poulet and Petersen, 2008), but also reduces the sensitivity of the cortex to passive whisker deflections (Crochet and Petersen, 2006; Crochet et al., 2011). However, interestingly, robust responses are maintained for active contact events when the whisker collides with an object placed in the whisk field (Crochet and Petersen, 2006; Crochet et al., 2011). Furthermore, in an active touch condition where the whisker repeatedly collides with an object both intra-neuronal correlations and low frequency power of membrane potential recovers close to their passive state values (Crochet et al., 2011). We theoretically investigated whether these phenomena could be explained by the hypothesis that vibrissa dynamics mediates negative CLE feedback to the brain, influencing cortical dynamics in a way analogous to fish swimming behavior. Notably, while multiple different mechanisms are involved in brain state transition (Goard and Dan, 2009; Polack et al., 2012; Pinto et al., 2013; Fu et al., 2014; Poulet et al., 2012; Zagha et al., 2014; Schneider et al., 2014), for which sensory input is not always necessary (Poulet and Petersen, 2008; Poulet et al., 2012), this does not exclude the role of CLE feedback under physiological conditions. Indeed, although brain state transitions can happen at whisking onset even after the sensory nerve is cut (Poulet and Petersen, 2008), our analyses revealed that the latency of brain state transition, measured by whisking-related reduction in low frequency power of cortical LFP or in thalamic spiking rate, was delayed under this condition as compared to the control condition (Supplementary Section S2). This supports a physiological role for sensory input in the brain state transition.

Here, we construct a simple model of the vibrissa system to investigate the possible role of CLE interaction. The model comprises excitatory and inhibitory cortical populations dynamically interacting with a single vibrissa (Figure 4A; see Methods for details). In our model, synchronous membrane potential fluctuations arise endogenously from interplay between the build up of excitatory cortical activity by recurrent connections and their eventual suppression by adaptation after ca. 1 s in each cell (see Methods). To model the brain/body interaction, we assumed that cortical excitatory neurons drive whisker motor behavior and both excitatory and inhibitory neurons receive the sensory input that reflects whisker angle. Based on our theory, Figure 1, we hypothesized that synchronous neural fluctuations could be reduced during whisking by negative CLE feedback. We modeled this by assuming that cortical excitatory neurons are activated by whisker retraction but this activation drives whisker protraction (see Figure 4A). (Note the specific biological implementation of the negative CLE feedback is not central to our claims, and there are several other plausible schemes, see Discussion and Supplementary Section 6.) We also model a central pattern generator (CPG, likely located in the brain stem) (Hill et. al, 2011) that, once turned on, rapidly cycles the position of the vibrissa back and forth at around 10Hz.

Sensory input during contact events involves both contact-related and whisker angle-related signals (Diamond et al., 2008). To capture this sensory input we modeled a single whisker as two stiff mass-less sections connected by hinges at the center and the base, which are constrained by muscles (simple torsion springs). Whisking is implemented by driving the equilibrium position of the base spring. The center spring has an equilibrium value of zero angular displacement and thus tends to align both sections (see Supplemental Information Section 3). We start with a simple non-bending stiff whisker but later vary the stiffness parameter (see Methods). A horizontal solid wall is placed above the whisker and, as the whisker collides with the wall, the whisker tip stops, see Figure 4A, interrupting CLE feedback about whisker angle. Whisker contact not only involves a modification of CLE feedback but also invokes exafferent input, or contact-detection signal, that results from the stereotypical response of pressure sensitive cells in the trigeminal ganglion (Szwed et al., 2003). We model this as a brief square wave pulses (ca. 25 ms) triggered by each contact event (see Methods).

Our key hypothesis, i.e., that whisker position mediates negative CLE feedback to cortical neurons, allowed us to reproduce rodent brain state findings that are observed under three behavioral conditions (the quiet condition, the whisking condition, and the active touch condition; Figure 4B [Crochet et al., 2011]). Specifically, the synchronous low frequency membrane potential fluctuations generated during the quiet condition (open-loop: the whisker position is fixed at 0) were significantly suppressed during the whisking condition (closed-loop: the whisker is driven by the CPG and cortex) (Figure 4C). In effect, the negative CLE feedback reduced the gain of the cortical system, see conceptual model Figure 1, and replaced prominent (ca. 1 Hz) synchronous fluctuations of the membrane potential with fast (ca. 10 Hz), but weak, fluctuations locked into the whisking cycle. As a result, the average inter-neuronal correlation of membrane potential for pairs of neurons was also suppressed (Figure 4D).

During the active touch condition, occasional whisker touch events stopped the whisker tip on the object's surface (Figure 4B; touch events marked by red bars). Both the low frequency power and cross-correlation of membrane potential fluctuations were partially recovered during the active touch condition in agreement with experimental results (Crochet et al., 2011) (Figure 4C and D, red lines). We hypothesized that this recovery was a consequence of intermittent interruption of negative CLE feedback during whisker contact events. To verify this, we examined whether this result depended on the stiffness of the whisker. We also hypothesized that, if the whisker is very flexible, then the base angle would change continuously, despite contact of the tip, fully preserving CLE feedback. In contrast, a very stiff whisker would come to complete rest on the wall during each contact maximally interrupting CLE feedback (see also Supplementary Figure S3B). We found that the recovery of 1-5 Hz power was indeed predominantly caused by the interruption of CLE feedback and not touch-evoked exafferent input (see Supplementary Figure S3C) demonstrating that these transient interruptions of CLE interaction can rapidly switch the active brain state to a passive brain state.

This model also accounted for observed brain state dependent changes in sensory processing (Crochet et al., 2011) without assuming additional mechanisms (Lee et al., 2008; Nguyen and Kleinfeld, 2005). In agreement with experimental results we found that passive exafferent input evoked large sensory responses in the quiet condition but markedly smaller responses during the whisking condition (Figure 4E) (Crochet et al., 2011). Again, this was because negative CLE feedback decreases the gain of the cortical circuit (Supplemental Section 4). However, in agreement with experimental data, cortical neurons exhibited more reliable responses to active touch events (Figure 4B and E). This was because the negative CLE feedback, that suppressed neural fluctuations during whisking, was transiently removed during active touch events allowing endogenous brain dynamics to amplify the cortical response to exafferent input (Figure 4E and F). Consequently, active touch events combined large sensory evoked responses (signal) and low background fluctuations (noise).

To quantify this effect, we computed a discriminability index (see Methods) that measures the separation between the distributions of membrane potentials in the presence or absence of sensory events. The value of the index was similar for perturbations in both quiet and whisking conditions, i.e., although the deflection-evoked response (signal) was greater in the quiet condition, so were background fluctuations of membrane potential (noise) (see Figure 4E and F). In contrast, the discriminability index was greater for active touch events (Figure 4E and F). Hence, the model suggests that cortical neurons are selectively sensitive to interruption of the animal's own active sensing. All these results are intuitively reproduced in a simplified model of CLE interaction, demonstrating the robustness of our findings (Supplemental Section 4).

Neurofeedback modulates brain dynamics and behavior

We further tested a core prediction of our theory by examining the impact of an artificially constructed CLE feedback on brain dynamics. Specifically, we investigated whether we could use an ECoG-based fast neurofeedback technique to modulate brain dynamics and subsequently behavioral performance in primates. This setup allowed us to observe the impact of CLE interaction without involving efference copy signal (which can induce brain state transition; Zagha et al., 2014; Schneider et al., 2014) and to investigate the possible clinical relevance of CLE feedback. We implemented real-time neurofeedback between the visual areas (occipitotemporal and parietal cortices) of a fixating macaque monkey (equipped with a 128-channel ECoG on the cortical surface, Supplementary Figure S5A) and a visual stimulus (equivalent to the environment in the fish study or vibrissa feedback), while the monkey was fixating (Figure 5A). During the presentation of a visual grating stimulus (Figure 5A), we decoded orientation (vertical or horizontal grating), in real-time, from ECoG activity with a support-vector machine. During neurofeedback sessions, a computer monitored the output of the classifier (the decision value), a nonlinear projection of the ECoG activity, and then modified the stimulus presented on the screen in real-time. A large and positive (or negative) decision value indicated that ECoG activity was likely driven by the horizontal (or vertical) grating stimulus, respectively. When the decision value exceeded a positive threshold, the computer showed a vertical grating for the subsequent 100 ms period; vice versa, exceeding a negative threshold triggered the presentation of a horizontal grating (see Methods). If the decision value fell between these two thresholds, a grey screen was presented. This protocol effectively approximates *negative feedback* (the closed-loop condition) on the decision value dynamics of the animal (however see Discussion). To distinguish between the influence of neurofeedback from the visual stimuli alone, we used a control condition in which the monkey was presented with a recording of the visual stimuli that emerged during the closed-loop condition (a replay condition). Thus, as in our conceptual mode (Figure 1A), and analogous to the fish experiment (see Figure 2A), corresponding closed-loop and replay conditions share identical visual stimuli and only differed by the presence of neurofeedback. Again, this allowed us to examine the exact effect of CLE feedback, rather than sensory input on brain activity and behavior.

We observed that neurofeedback immediately altered the dynamics of the decision value and thus brain dynamics, without requiring any prior training or habituation of animals. The power spectrum of the decision value below 5 Hz significantly differed under the two conditions (Figure 5B). While the 0-1 and 3-4 Hz (summarized in Figure 5C) frequency ranges were suppressed, the 1-3 Hz frequency range was enhanced during the closed-loop condition relative to the replay condition. This oscillatory modulation of the power spectrum could have arisen from delay (ca. 100 ms) in negative CLE feedback, but it would also depend on various factors, including the kinetics of the decision value, and early visual stream processing. However, the exact mechanisms underlying the changes in the power spectrum are not the focus of this study. We found that not only were the dynamics of the decision value altered by neurofeedback, but also that there was significant improvement in the monkey's fixation performance (Figure 5D). Specifically, the distance of the eye position from the fixation point was quantified and compared under the two conditions. The fixation performance at 1600-2100 ms after the appearance of the fixation point (about 300-900 ms after typical grating onset; Supplementary Figure S5C) in the closed-loop condition was enhanced (Figure 5E). To establish a quantitative relationship between changes in neural dynamics and behavior, we analyzed the trial-to-trial correlation between the power spectrum of the decision value and fixation performance. The reduction of fluctuation in the decision value at specific frequencies correlated with improved fixation performance (Figure 5F). In particular, the neurofeedback-induced reduction of decision value fluctuations in the 3-4 Hz frequency range (estimated from data before the behavioral improvement; see Supplementary Figure S5D) showed a significant correlation with improved fixation performance (Figure 5G). Taken together, these results indicate that in primates artificial neurofeedback can change brain dynamics related to sensing and task-related behavior, which was consistent with our fish and rodent studies. Although the causal chain of events underlying these changes may also involve internal brain mechanisms for state transitions (e.g., neuromodulators; see Discussion), we assert that neurofeedback constitutes the primary cause of these changes as it is the only difference between each condition.

**Discussion**

Brain state transitions triggered by the onset of active interactions between the brain and environment, represent a major neuronal mechanism shaping sensing and behavior. In this study, we use both theory and experiment to support the idea that negative CLE feedback inhibits network gain, which in turn, suppresses synchronous neuronal fluctuations and sharpens sensory responses. We generalize and support the theoretical framework in three diverse animal model systems summarized in Figure 6. In each system, we show that negative CLE feedback regulated real-time brain state and animal behavior. Specifically, CLE feedback quantitatively predicted cell-specific suppression of neural oscillations in zebrafish, enhanced signal-to-noise ratio for active sensing in rodent, and enhanced task performance in primate vision.

The importance of using naturalistic sensory stimuli to study and manipulate brain state dynamics is widely demonstrated (Felsen and Dan, 2005). However, an important prediction of our theory (Figure 1, Supplemental Information Section 7), supported by our experimental findings (Figures 2, 3, and 5), is that brain dynamics during active sensing cannot be fully recapitulated or re-encoded, even if the same sensory input is precisely recorded and replayed back into a passive brain. These results provide evidence that brain state neurometrics and behavioral psychometrics during active behaviors can only be accurately understood by a quantitative account of ongoing brain-environment interactions (O'Regan

and Noë, 2001).

**Neuronal gain control by CLE feedback**

The formal component of our theory, i.e., that CLE feedback can modulate a system's gain, is well documented in dynamical systems theory and control theory (e.g., Aströmand Murray, 2010). This gain control occurs even though the instantaneous effect of the pathways mediating feedback is purely additive (c.f. Figure 1) because the effect of repeated cycles of feedback accumulates over time. For example, input $I$ to a linear dynamical system with feedback strength $b$ makes a first direct contribution $I$ to its response, but then also makes subsequent contributions as this initial response cycles around a feedback loop—a contribution of $bI$ after one cycle, $b^2 I$ after two cycles and so on. The cumulative sum of these contributions $I + bI + b^2 I + \ldots = I/(1-b)$ is equivalent to divisively scaling the input magnitude by a factor that depends on the feedback strength, i.e., effectively changing the system's gain. Thus, a constitutively active closed-loop feedback that mediates multiple action-perception cycles is essential for the form of gain control we propose.

This means that discrete and intermittent involvement of reafferent input does not imply gain modulation. For example, the classical reafference principle explains neuronal responses by a one-time detection of the mismatch between an efference copy (predicted) and reafferent (actual) (Von Holst, 1954). However, this situation is likely an inaccurate idealization to describe the closed-loop systems studied here. For example, in the zebrafish system, swim bouts typically occur every 700 ms and this interval closely overlapped with the peak of the estimated environmental feedback interaction (Figure 3A and Supplementary Figure S1C). Hence, the neural responses in the fish experiment suggest a more dynamic system, where neural activity evoked by many cycles of action and sensation are continuously and mutually interacting.

**Neuronal mechanisms of negative CLE feedback**

The presence of continuous negative CLE feedback during active behavior is fundamental for our theory. Although, as a higher order theory, it is agnostic to the detail of the neural implementation, we discuss below the implications for each system. In zebrafish, the presence of negative CLE feedback during fish swimming behavior would seem a priori necessary for optic-flow stabilization behavior because the fish must act in opposition to perceived optic flow in order to minimize horizontal displacement (Wolf et al. 1992, Fry et al., 2009). Interestingly, neurons that received strong negative CLE feedback and were substantially stabilized were located in the cerebellum (Figure 3C) consistent with a theoretical viewpoint that the cerebellum mediates this optomotor response by converting sensation into action; e.g., by completing the action-perception cycle (Ito 1972, Lisberger et al. 1987, Kawato 1999).

Our rodent whisker model explores the overall effects of negative CLE feedback mediated by the cortical-whisker circuit during active whisking. As in the fish study, the theoretical assumption of negative CLE feedback is consistent with the idea that the barrel cortex is involved in control. Specifically, the barrel cortex comprises a nested set of servo control loops that regulate various aspects of whisker dynamics (Ahissar and Kleinfeld, 2003). At the level of the whole vibrissa system, multiple parallel and nested feedback loops, both positive and negative (Ahissar and Kleinfeld, 2003) most likely exist. While our model is abstract in terms of the known complexities of vibrissa system anatomy and exact concordance of the model with vibrissa system anatomy is beyond the scope of this paper, we provide several possible schemes for experimentalists to examine in Supplemental Information Section 6.

The artificial CLE feedback in the primate experiment implements negative feedback, although this is less obvious than in our other studies and simple models. A system is defined to be under negative feedback if this feedback causes perturbations to the system's state to decay faster in time and suppresses fluctuations due to noise. Hence, neurofeedback in the monkey experiment is a hybrid of positive and negative feedback—the feedback is negative in the 0-1 Hz and 3-4 Hz frequency bands and positive in the 1-3 Hz frequency band, because the feedback reduces or increases fluctuations of the decision value in these frequency bands, respectively (Figure 5B). In this experiment, the system's state was quantified by the decoder's decision value, with positive and negative values indicating vertical-like and horizontal-like brain activity patterns, respectively. Due to experimental constraints of delays in neurofeedback processes and unidentified dynamic properties of the brain, imposing purely negative neurofeedback in this experiment was not possible. However, negative CLE feedback in the 3-4 Hz frequency band is strongly correlated with enhanced fixation performance (Figure 5F), supporting its behavioral significance.

**Comparison to internal mechanisms of brain state transitions**

The proposed mechanism of brain state transitions by CLE feedback is qualitatively different from the many other mechanisms governing brain state that operate inside the brain (Goard and Dan, 2009; Polack et al. 2012; Pinto et al., 2013; Poulet et al., 2012; Zagha et al., 2014). Specifically, the CLE feedback mechanism involves the dynamic coordination between brain activity and body/environmental dynamics (c.f. Figure 3D), a continuous reciprocal interaction that is critical for many forms of goal-directed behavior. In this regard, our theory directly links a mechanism of brain state transitions to behavior and is closed from an explanatory view, i.e., without assumptions about upstream causes.

In our zebrafish study, an alternative interpretation of our findings is that correlated neural fluctuations are suppressed directly by an efference copy signal (Zagha et al., 2014; Schneider et al. 2014) due to greater motor commands in the closed-loop condition (Supplementary Figure S2). However, in opposition to this view, we found that motor activity correlated positively with fluctuations in each cell, and were not associated with pairwise neuronal correlations (Supplementary Figure S2). Hence, we cannot account for the results by any known functions of efference copy signals. Moreover, the primate neurofeedback allowed us to examine the impact of CLE feedback in the absence of the overt involvement of the motor system, i.e., feedback was only dependent on readout from visual areas. This allows us to conclude that CLE feedback alone, in the absence of noticeable motor activity, is sufficient to modulate brain activity and behavior.

Importantly, brain state control by CLE feedback is not mutually exclusive with other mechanisms, such as thalamo-cortical input (Poulet et al., 2012) or neuromodulation (Goard and Dan, 2009; Polack et al. 2012; Pinto et al., 2014; Fu et al. 2014) because these may also be involved in mediating the action-perception cycle. Furthermore, brain state transitions also occur in the absence of CLE feedback, such as open-loop behaviors (e.g., the onset of running that does not change the visual screen or grooming) (Niell and Stryker, 2010, Fee et al. 1997), during sleep (Vyazovskiy et. al 2011; Steriade, 2005), or under dissection of the sensory nerve (Poulet and Petersen, 2008; Poulet et al., 2012). Mechanisms underlying brain state transitions are likely to be highly redundant even in the absence of essential mechanisms, such as thalamo-cortical input (Poulet et al., 2012) or corollary discharge (Schneider et al. 2014), albeit involving further delay (see also Supplementary Section 2). Such functional redundancy may help to maintain the stability of brain state (Fu et al. 2014; McGinley et. al, 2015; Otazu et al. 2009; although with exceptions, see Vyazovskiy et. al 2011). Furthermore, the relative importance of internal and external mechanisms might adaptively change in an experience-dependent manner (Nachestedt et. al 2014).

**Reafferent feedback in active sensing and revision of the reafference principle**

In the whisking model, we proposed that the regulation of cortical gain by CLE feedback could explain enhanced active touch. Specifically, negative CLE feedback during whisking reproduces suppressed fluctuations and reduces responses to passive whisker stimulation (see Figure 4F). However, robust neuronal response to active touch events could be explained by the interruption of CLE feedback when the whisker is driven into an external object. Such events interrupt CLE feedback, transiently releasing the cortex from a low gain state and enhancing sensory responses to salient sensory stimuli (Figure 4E, Supplemental Information Section 4). This mechanism for active touch contrasts with the account of sensory processing suggested by the reafference principle (Von Holst, 1954), which postulates that motor efference is discounted from sensory input, allowing animals to sense exafferent signals (externally caused sensory input) without being confounded by the consequences of their own motor actions. In contrast, our theory suggests that the sensory system is insensitive to pure exafference during active sensing (Figure 4F), but sensitive to the interruption of reafference which may allow animals to focus attention on the consequences of their own motor actions.

While it is straightforward to generalize this idea to other tactile systems, its implication for other sensory modalities is less clear. However, in theory, CLE feedback could be interrupted anywhere along the action-perception cycle, thus dynamically regulating neuronal gain. The timely interruption of this feedback could serve as a general mechanism for temporarily accentuating neuronal responses against a background of reduced noise (Scott, 2004, Hafed et al., 2011). For example, cerebellum neurons, which are strongly involved in the sensory-motor cycle, could be suppressed in anticipation of salient sensory events by a relevant brain area, such as the reticular formation (Kinomura et al. 1996).

**Rapid neurofeedback and virtual reality-based behavioral enhancement**

The primate findings demonstrate that neurofeedback can modulate brain dynamics (Figure 5B and C) and enhance task-related behavior (Figure 5D, E, F, and G). These results portend the use of CLE feedback as an interventional tool for behavioral enhancement in closed-loop feedback therapies. Unlike more conventional neurofeedback techniques (Zoefel et al., 2011, Shibata et al., 2011) that require training periods to affect behavior, our technique enhanced fixation performance within seconds and required little supervision. This rapid neurofeedback was achieved by crucial differences in the design between this technique and conventional neurofeedback. Conventional protocols require conscious/unconscious human learning that is supported by an appropriate neurofeedback signal. In contrast, our protocol directly intervened on the fast time scale of the ongoing sensory stream. Our approach has the potential to be more stable, with more rapid, user-independent, and effective behavioral control than current conventional decoded neurofeedback methods.

The neurophysiological reason underlying the improved fixation performance is not clear. However, the high-contrast grating stimulus likely serves as a distractor for the monkey and makes fixation more difficult. During the neurofeedback protocol, the 3-4 Hz frequency component of this distractor signal was suppressed by the feedback, possibly inhibiting the formation of a grating perception. If so, it is possible that such CLE feedback may help to allocate attention in physiological settings. Specifically, our model predicts that, animals may show less attention to sensory modalities that provide ongoing negative CLE feedback. A full investigation of these issues will require further work.

These findings emphasize the importance of the development of natural and virtual reality systems that provide well-controlled quantitative measurements of CLE interactions. Such novel systems would be useful for studying how healthy and disease brain physiology emerges from real-time brain-body-environment interactions and may suggest new methods for manipulating or enhancing brain physiology and behavior with neurofeedback technology.

In sum, our findings suggest that context-dependent brain function and flexible behavior may only emerge from context specific CLE interactions. This supports the idea that CLE interactions may have a broad impact on our cognition (Clark, 2008) and may also shape social interactions (Froese et al., 2014).

**Materials and Methods**

**Simple Conceptual Model:** The time traces in Figure 1 were calculated by Euler's method, integrating the equations presented in the text with $w = -10$, $\tau = 1$, and time step $dt = 0.01$.

**Zebrafish experiment:** We analyzed the calcium signal (ΔF/F) at various sample frequencies (ca. 2–3 Hz) across 1908 cells in 32 fish, see Ahrens et al. (2012) for full description of experimental method. In addition to calcium sources (putative neurons), swim vigor was measured by taking electrical recordings from motor neurons in the fish's tail. In the original study, the gain (i.e., the multiplicative factor between fictive swim power, and the speed of visual feedback) was alternated between a high and low gain condition every 30 s.

We studied cells analyzed in Ahrens et al. (2012), which showed modulations of the mean ∆F/F, depending on the gain conditions. This gain alternating protocol is not relevant to the current study. To reduce this variability in data, we subtracted the mean activity level in each gain setting in our analysis (from both brain and behavior variables). Notably, our main results were qualitatively the same, even without such subtraction of the means.

We analyzed data taken from a 6-min recording of 1–6 prominent calcium sources per fish, putative neurons, across 600 trials. In the first 3 min, the fish performed the closed-loop optomotor behavior. For the subsequent 3 min, each fish was presented with an open-loop stimulus (no longer actively controlling their environment), which is a repeat of what the animal experienced in the previous 3 min. In this period, sensory input to the animals was identical and the only difference was the absence of closed-loop dynamics, allowing us to study how reafferent signal affected neural activity.

In Figure 2B, low frequency power was calculated as the mean log power of a neuron in the range of 0.01–0.15 Hz, averaged over all recorded neurons. Correlation was calculated as the average pairwise correlation (6589 pairs of cells were analyzed) between neurons in the same fish, averaged over all trials. Figure 2C plots the change in this pairwise correlation against the change in the log low frequency power (averaged for the pair) between the closed-loop and the replay condition. Note: for fish data, non-parametric tests were used and we did not assume normality. In addition, both results reported in Figure 2B were highly significant under a paired *t*-test, with p values $<10^{-5}$.

We fitted the data with linear filters that describe the interaction between individual neurons and the environmental variable. Note: here we define the environmental variable (E) as the activity of motor neurons, whose history uniquely determines the visual stimulus. We calculated the linear filter that minimizes the mean square error between a driven variable $y(t)$ and the convolution of a driving variable $x(t)$ and filter $F(t)$ over time. Filters were constructed as a superposition of Laguere functions. We use Laguere functions up to the order that stopped the optimization process by choosing the best Akaike Information Criterion (Akaike, 1987). Almost all filters had an order that was mid-range between 1 and 15, indicating the robustness of the chosen order.

We inferred two filters for each cell. We determined how the environment drives the brain, $F(t)$, by directly calculating the filter between the closed-loop environment and brain in the replay condition $E \rightarrow B'$, see Figure 3A (magenta line). We determined how the brain drives the environment, $G(t)$, by first calculating the residual variability of brain in the open-loop condition that cannot be accounted for by the closed-loop environment, i.e., $R_{B'}(t) = B'(t) - F * E(t)$ and subsequently calculating how $R_{B'}(t)$ drives the environment in the open-loop condition, effectively determining $B' \rightarrow E'$. Self-feedback is then straightforwardly estimated as the convolution of both filters $H(t) = F * G(t)$.

In Figure 2B, we calculated the low frequency power as the mean log power of a neuron in the range (0.01–0.15 Hz) and took the ratio between the closed-loop and replay condition. To analytically calculate this low frequency power ratio from the estimated filters, we can first write the dynamics of the brain in the closed- and open-loop conditions in the frequency domain as

$$B(\omega) = H(\omega)B(\omega) + \epsilon(\omega) = (1 - H(\omega))^{-1} \epsilon(\omega)$$

$$B'(\omega) = H(\omega) B(\omega) + \epsilon'(\omega),$$

where $\epsilon(\omega)$ and $\epsilon'(\omega)$ are the Fourier transforms of the noise in the closed- and open-loop conditions that we assume as having the same frequency spectrum. The ratio of the power between each condition is simply

$$\frac{|B(\omega)|^2}{|B'(\omega)|^2} = \frac{1}{|H(\omega)|^2 + |1 - H(\omega)|^2},$$

where $H(\omega)$ is the estimated combined filter in the frequency domain. From this curve we can then straightforwardly calculate the mean log power of a neuron in the range (0.01–0.15 Hz).

In Figure 3D, we calculated the $E \to B'$ filter in the replay condition and the $E \to B$ filter in the closed-loop condition as above, but using the Hermite rather than the Laguere functions to capture the acausal ($t$<0) side of the filter. Notably, the $E \to B$ filter in the closed-loop condition generally has an acausal component, because the brain $B$ and the environment $E$ are mutually interacting (see below). On the other hand, the $E \to B'$ filter in the replay condition is identical to the $F(t)$ filter defined in the previous paragraph (although with minor differences due to the use of the Hermite- rather than Laguere-based functions). Hence, the two filters ($E \to B'$ and $E \to B$) are generally different and this difference originates from the presence of CLE feedback ($B \to E$ interaction) in the closed-loop condition. Here, we quantify this difference using a simple linear model. In the closed-loop condition, the system is described by

$$B(\omega) = F(\omega)E(\omega) + R_B(\omega)$$

$$E(\omega) = G(\omega)B(\omega) + R_E(\omega)$$

in the Fourier domain. Note that we again assumed the same interactions $F(\omega)$ and $G(\omega)$ in the closed-loop and replay conditions. Based on the first equation, the second equation is also described as $E(\omega) = (1 - H(\omega))^{-1}(G(\omega)R_B(\omega) + R_E(\omega))$. Therefore, the $E \to B$ filter in the closed-loop condition is

$$\frac{B(\omega)E^*(\omega)}{E(\omega)E^*(\omega)} = F(\omega) + \left(\frac{E(\omega)R_B^*(\omega)}{E(\omega)E^*(\omega)}\right)^* = F(\omega) + \left(\frac{G(\omega)}{1 - H(\omega)}\right)^* \frac{|R_B(\omega)|^2}{|E(\omega)|^2}$$

where * describes complex conjugate. Hence, this filter is different from $F(\omega)$ by the second term. To predict the second term in the absence of knowing $B$, we assume $|R_B(\omega)|^2 \approx |R_{B'}(\omega)|^2$, where the latter spectrum is based on the residual $R_{B'}$ computed in the replay condition.

**Whisker model:** We model a whisker comprised of two sections that are connected by hinges at the center and the base, which are constrained by muscles (simple torsion springs), see Supplementary Figure 3A. Whisking is implemented by driving the equilibrium position of the base spring. The center spring has an equilibrium value of zero angular displacement and thus tends to align both sections. A horizontal solid wall is placed above the whisker and, as the whisker collides with the wall, it deforms accordingly (Supplementary Figure 3B). By adjusting the relative stiffness of each torsion spring, we can control the degree to which the base angle is affected by contact events, e.g., if the whisker is very flexible, the base angle will change continuously, despite contact of the tip, see Supplementary Figure 3B. Note: the feedback in this model could be equally mediated by velocity or even acceleration, rather than whisker position, but neither choice would make a significant difference to our conclusions, as long as the CLE feedback constitutes a net negative feedback. We fixed the base spring, k1 = 1, and identified the stiffness of the whisker with k2. For Figures 4A–E, we considered a rigid whisker, but relaxed this assumption in Figure 4F and Figures 3B and C. We simulated the dynamics of the whisker by minimizing a Lagrangian description of the configurational energy of the massless whisker under the constraint of the solid wall.

We model a cortical circuit comprising $N$ excitatory and $N$ inhibitory neurons that interact with a single whisker. Dynamic activity of neuron $i$ ($i = 1,\ldots,N$ are excitatory and $i = N+1,\ldots,2N$ are inhibitory, with $N = 100$) is modeled as a linear dynamical system by

$\dot{x}_i = -x_i + \sum_{j=1}^{2N} \omega_{ij} x_j - a_i - \omega_{\theta y} \theta + \xi_i + I$

Differential equations are solved by the forward Euler integration method with time-bin $dt = 0.5$ ms. Hereafter, all time derivatives are taken to represent single-step differences divided by $dt$ (e.g. $\dot{x}(t) = [x(t+dt) - x(t)]/dt$), but we omit the ms time unit. $\omega_{ij}$ is the synaptic strength from neuron $j$ to $i$, $a_i$ is an adaptation current that produces low frequency (ca. 1 Hz) up/down-like oscillations (Compte et al., 2003; Gentet et al., 2010; Curto et al., 2009) in the absence of neuron/whisker interactions, $\theta$ is whisker angle (positive: protracted and negative: retracted) interacting with neurons with weight $\omega_{\theta y} = 0.002$, $I$ is exafferent input that takes a non-zero value upon whisker stimulation, and $\xi_i$ is independent white noise of unit variance added to each neuron. Here, we interpret $x_i$ as both the firing rate and membrane potential, assuming a roughly linear relationship between the two. Entries in the connectivity matrix are assigned as $\omega_{ij} = b_{ij}J + b'_{ij}g$ for excitatory synapses ($j = 1,\ldots,N$) and $\omega_{ij} = -b''_{ij}g$ for inhibitory synapses ($j = N+1,\ldots,2N$), where $b_{ij}, b'_{ij}, b''_{ij}$ are all random binary values that take $b_0$ with probability $p = 0.1$ and 0 with probability $1-p$, respectively. The weights are scaled by $J = \frac{1}{pN}$ and $g = \frac{g_0}{\sqrt{2Np(1-p)}}$, so that dynamics are insensitive to the parameter values of $p$ and $N$. Note: all excitatory and inhibitory neurons behave similarly but adding sparse recurrent connections between randomly selected pairs neurons can account for inter-neural variability (Renart et. al 2010). The parameter $g_0 = 0.05$ controls inter-neural variability, and a value less than 1 reproduces highly synchronized up/down-like fluctuations during the quiet state. Finally, the scaling of the connectivity matrix $b_0$ is determined such that the lead eigenvalue of the connectivity matrix is close to unity ($\approx 0.975$) and the dynamics are close to instability. The adaptation current is integrated as

$$\dot{a}_i = -0.07\, a_i + 0.008 x_i$$

Over time, the adaptation variable slowly builds upon neural activity and suppresses neurons, resulting in the ca. 1-Hz oscillation. Consequently, in the absence of whisking, implemented by setting $\omega_{\theta y} = 0$, this simple network reproduces the power spectrum and cross-correlogram of neurons in the cortex (Figure 4).

The whisker protraction of the base ($\theta_1$, see Supplementary Figure 3A) is driven by the sum of activity in the excitatory population and an external CPG activity, $u$; i.e., it is modeled as

$$\dot{\theta}_1 = -0.93\, \theta_1 + \frac{\omega_{y\theta}}{N} \sum_{i=1}^{N} x_i + u$$

and is driven by the sum of activity of the excitatory population. $\omega_{y\theta} = 0.085$ describes the relative strength of the cortex vs. the CPG in driving the whisker variable. With this parameter, the whisker is mainly driven by the CPG and is modulated by cortical activity. During whisking, the whisker rhythm is imposed by the CPG. The rhythm is generated by a simple stochastic oscillator, given by

$$\dot{u} = -.98u + 2\pi F_{whisk} v + \xi_u$$

$$\dot{v} = -.98v - 2\pi F_{whisk} u + \xi_v,$$

where $F_{whisk} = 10$ Hz is the frequency of the oscillator and $\xi_u, \xi_v$ are independent Gaussian white noise.

Aside: In the current model, most excitatory neurons respond to whisker retraction and drive whisker protraction. Adding a separate counterpart population that responds to whisker protraction and drives whisker retraction does not change the model's behavior. In the current model, the tuning of cortical neurons to whisker position (Diamond et. al 2008) is mainly inherited from thalamo–cortical input.

Passive whisker stimulations are simulated by injecting exafferent input $I = 0.035$ to the cortical neurons for 25 ms. The magnitude of the exafferent input is selected such that it approximately matches the evoked change over the standard deviation of the membrane potential ($\Delta Vm / \sigma(Vm)$) in response to magnetic whisker deflection during the whisking condition (Crochet and Petersen, 2006).

During active touch events, we injected exafferent input ($I = 0.035$) to the cortical neurons on contact, for the duration of the contact event, but for no longer than 25 ms. The model was run for 200 s in the closed loop, open-loop, and sustained period of active touch to calculate all quantitative measures. The power in Figure 4C is averaged over all neurons and a quadratic spline fitted to the data.

**Chernoff discrimination:** To calculate signal-to-noise-ratios, we calculated the Chernoff distance (Cover, 2012) between two probability distributions, $p_I(x)$ and $p_0(x)$, in the presence or absence of a sensory event, respectively. Specifically,

$$\Psi(p_I||p_0) \equiv -\min_{0<\lambda<1} \log \int p_I^{\lambda}(x) \, p_0^{1-\lambda}(x) \, dx$$

For our model, the probability distribution for each condition is well described by a Gaussian distribution

$$p(x) = |2\pi C|^{-1/2} \exp(-\frac{1}{2}(x-\mu)C^{-1}(x-\mu)),$$

where C and $\mu$ are covariance matrix and vector of means, respectively. By substituting this into the expression for Chernoff distance and employing the Gaussian integral identity and expressing the Chernoff distance in terms of $C_0, C_I$ and $\mu_0, \mu_I$, we calculate the covariance and mean between three neurons, randomly selected from the network described in the first section. We calculate covariances across ensembles of 500 networks every 10 ms for a period of 1 s, starting at the onset of the sensory event. Minimization with respect to $\lambda$ is computed numerically.

**Monkey experiment:** ECoG activity was recorded from 128 subdural electrodes implanted in a macaque monkey. All experimental and surgical procedures were performed in accordance with experimental protocols (No. H24-2-203(4)) approved by the RIKEN ethics committee and the recommendations of the Weatherall report, "The use of non-human primates in research".

Subject:

One monkey (male *Macaca mulatta*, aged 12 years, wild-type) was used in the experiment after magnetic resonance imaging. Before the implantation of subdural ECoG electrodes, the monkey was familiarized with the experimental settings and trained with a fixation task. During the fixation task, the monkey sat in a primate chair with the head in a fixed position using a helmet custom-made for the monkey. The monkey was housed individually in the room with a 12-h light–dark cycle (lights on at 8:00 AM), and participated in the experiment in the daytime. The monkey had also participated in other experiments involving fixation and voluntary saccadic eye movements more than half a year previous to this study.

Electrode Implantation:

Subdural electrodes were surgically implanted after the completion of fixation training. To anesthetize the monkey, we administered ketamine (5 mg/kg, intramuscular), atropine (0.05 mg/kg), and pentobarbital (20 mg/kg, intravenous). We adjusted the dose of pentobarbital based on monkey's response to pain and heart rate. We chronically implanted a customized 128-channel ECoG electrode array in the subdural space (Unique Medical, Tokyo, Japan; Nagasaka et al., 2011).

The monkey was rewarded only in trials for which fixation was maintained from FP onset until its disappearance. ECoG signals were recorded at a sampling rate of 1 kHz using a Cerebus data acquisition system (Blackrock, UT, USA) and down-sampled to 200 Hz. Behavioral variables, such as pupil size and eye position, were recorded with a custom eye-tracking system (Nagasaka et al., 2011). Visual stimuli were constructed in Pyschtoolbox and feedback implemented by streaming ECoG data in real-time into Matlab. We used LIBSVM (Chang and Lin, 2011) under Matlab to implement the SVM.

During a 10-min training session, the monkey passively viewed consecutive 500-ms blocks of either vertical or horizontal gratings in each trial (total of about 200 trials). We found that the stimulus was most reliably decoded by applying SVM to inter-electrode correlations. We collected 100-ms runs of ECoG data from 50 electrodes for each stimulus condition and constructed the SVM vector by calculating the correlation between electrodes and flattening half (upper right triangle) of the resulting matrix to form a vector. We trained an SVM to classify vertical and horizontal grating on all collected vectors.

Classification performance strongly depended on the time interval used for the classifier, typically reaching > 90% cross-validated classification accuracy with 400-ms-long intervals. In order to reduce temporal delay for neurofeedback, we classified stimuli using short 100-ms intervals. This typically gave > 60% cross-validated classification accuracy which is still markedly higher than chance level. A better classification performance was also achieved from grating orientations separated by 90, rather than 180 degrees, motivating our choice for horizontal and vertical gratings for the neurofeedback.

We then estimated the decision thresholds, by calculating a distribution of the decision values for each stimulus condition from the training data. These were roughly Gaussian with a positive and negative mean for the horizontal and vertical grating stimuli, respectively. We estimated the threshold for the decision value above ($\theta_{hor}$) and below ($\theta_{vert}$) which vertical and horizontal gratings are shown, respectively (see main text), by calculating the absolute value difference between the distributions and selecting the peaks of maximal difference. No gratings were presented for at least 1 s after the fixation point onset, to avoid presenting stimulus before the monkey fixated his eyes.

During testing, we consecutively sampled 100-ms runs of ECoG data, constructed an SVM vector, and used the SVM kernel to calculate the decision variable. The Matlab processing time for classifications was shorter than 100 ms. We were able to record 50−108 trials for each of the closed-loop and replay conditions per experiment. We present data from four recording days in the text. We recorded a minimum of 326 trials each under the closed-loop and replay conditions. Together with the calibration of the decoder, this added to a reasonable experimental length for the subject to maintain his attention. In addition, the number of trials was comparable to that of typical experiments performed in visual psychophysics.

The power spectrum of the decision value (Figure 5B) was calculated from the ECoG signal within 1000−2800 ms after the fixation point onset. Trials in which the monkey failed to fixate his eyes for at least 2800 ms were eliminated from the data set. The improvement in the performance was measured by the difference between the positions in each condition (Replay - Closed) relative to the average position (Replay + Closed) / 2 within each time-bin.

Normality for applying the *t*-test (Figure 5C) was examined using the Kolmogorov–Smirnov goodness-of-fit hypothesis test ($P > 0.07$).

The performance improvement for each trial pair was summarized by root-mean-square of this quantity measured over an interval of 1600–2100 ms. We computed the trial-to-trial correlation between the fixation performance improvement and the power spectrum of the decision value dynamics in the interval of 1000–2100 ms (Figures 5F and 5G): the power spectrum within this time period reflects the brain dynamics that precede the behavioral change. We also confirmed that the overall trend of the power spectrum of the decision value within this period (Supplementary Figure 5D) had not changed from that derived from the whole stimulation period shown in Figure 5B.

**Acknowledgements:** We would like to thank the authors of Ahrens et. al. (2012) for kindly providing the zebrafish data and Poulet et. al. (2012) providing the rodent whisking data. We would like to thank Charles Yokoyama and Alexandra V Terashima for comments on the manuscript, and Misha Ahrens, Hideaki Shimazaki, Hokto Kazama, Masanori Murayama, Sylvain Crochet and L.F. Abbott for helpful discussions.

**Author Contributions:** C.L.B and T.T. developed the theory. C.L.B. simulated the whisker model and analyzed the zebrafish data, in discussion with T.T. C.L.B. and T.T. planned the monkey experiment. C.L.B., S.T., T.Y., K.T. and Y. N. conducted the monkey experiment with the help of N.F. S.T. analyzed the monkey data based on the initial account by C.L.B. C.L.B. and T.T. wrote the manuscript.

Wolpert, D. M., & Ghahramani, Z. (2000). Computational principles of movement neuroscience. *nature neuroscience*, *3*, 1212-1217.

Y. Nagasaka, K. Shimoda, N. Fujii. Multidimensional recording (MDR) and data sharing: an ecological open research and educational platform for neuroscience. PloS one, 6(7), e22561 (2011).

Zagha, E., Casale, A. E., Sachdev, R. N., McGinley, M. J., & McCormick, D. A. (2013). Motor cortex feedback influences sensory processing by modulating network state. Neuron, 79(3), 567-578.

**Figures**

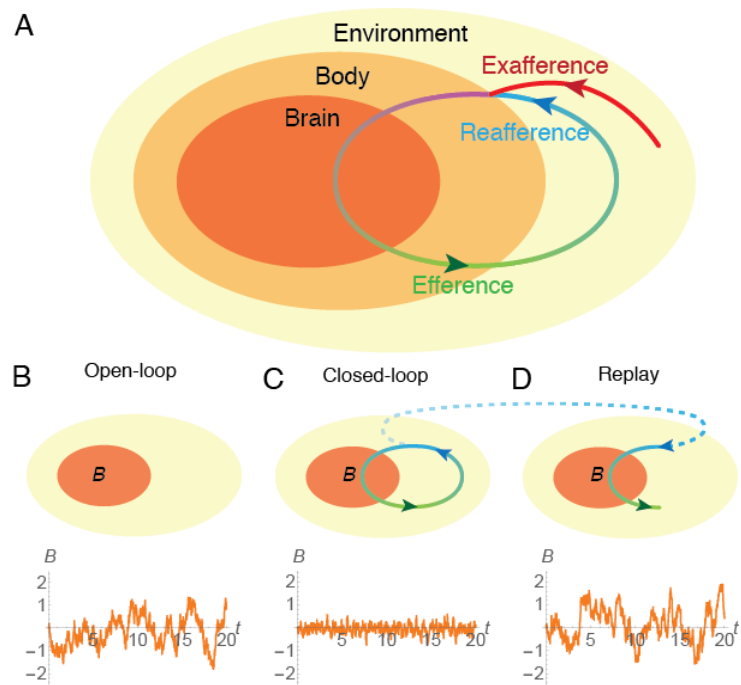

**Figure 1 |** A simple model of the brain-environment interaction. (A) A schematic description of brain-body-environment interactions during closed-loop behavior. The brain receives two types of sensory input: exafferent input that originates from the environment and reafferent input, which, while mediated by the environment, results from the consequences of an animal's own actions. (B-D) Schematic diagrams (Top) and the model's representative brain activity traces (Bottom) under the following three conditions. In the open-loop condition (B), the brain receives no reafferent input and exhibits collective activity that spontaneously fluctuates. In the closed-loop condition (C), reafferent input constitutes a CLE feedback to the brain. If this feedback is negative, the gain of the brain is reduced and fluctuations are suppressed. In the replay condition (D), the brain receives a replay of the reafferent input in

the closed-loop condition as exafferent input. Any differences from the closed-loop condition are caused by the absence of CLE feedback because the sensory input is identical to that in the closed-loop condition. In this condition, the gain of the brain is not suppressed and fluctuations are much larger than in the closed-loop condition with negative feedback.

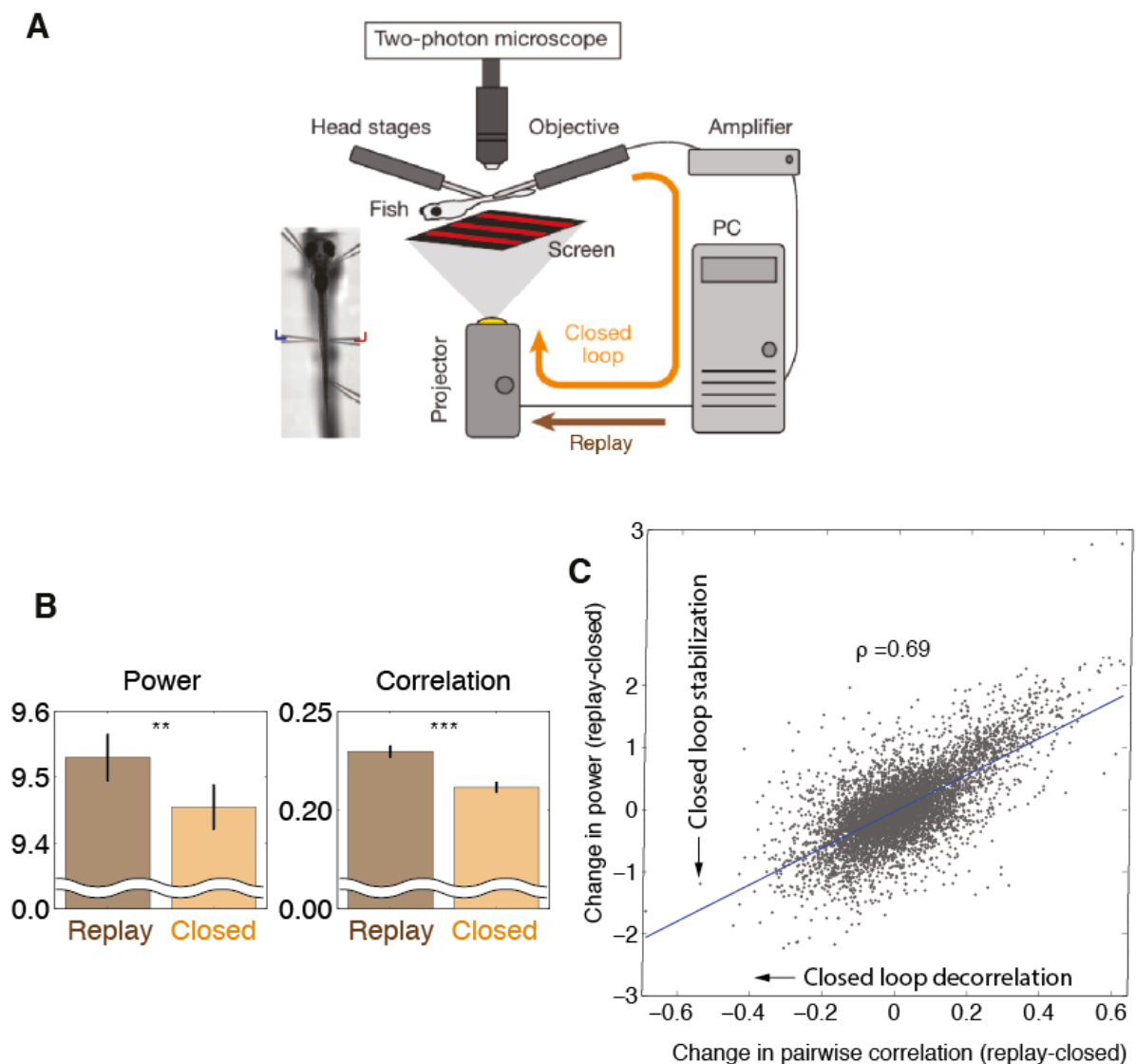

**Figure 2.** CLE feedback suppresses neural fluctuations and correlations.

(A) Photograph of a paralyzed larval zebrafish (left) in the experimental setup (right), supported by pipettes that record motor activity. (B) Population averages of logarithmic low frequency power (mean over interval of [0.01 0.15] Hz) (left) and pairwise intra-neural correlations (right) were both suppressed under the closed-loop condition relative to the replay condition. (C) These changes in pairwise correlations and low frequency power (replay – closed) were highly correlated in the recorded neurons.

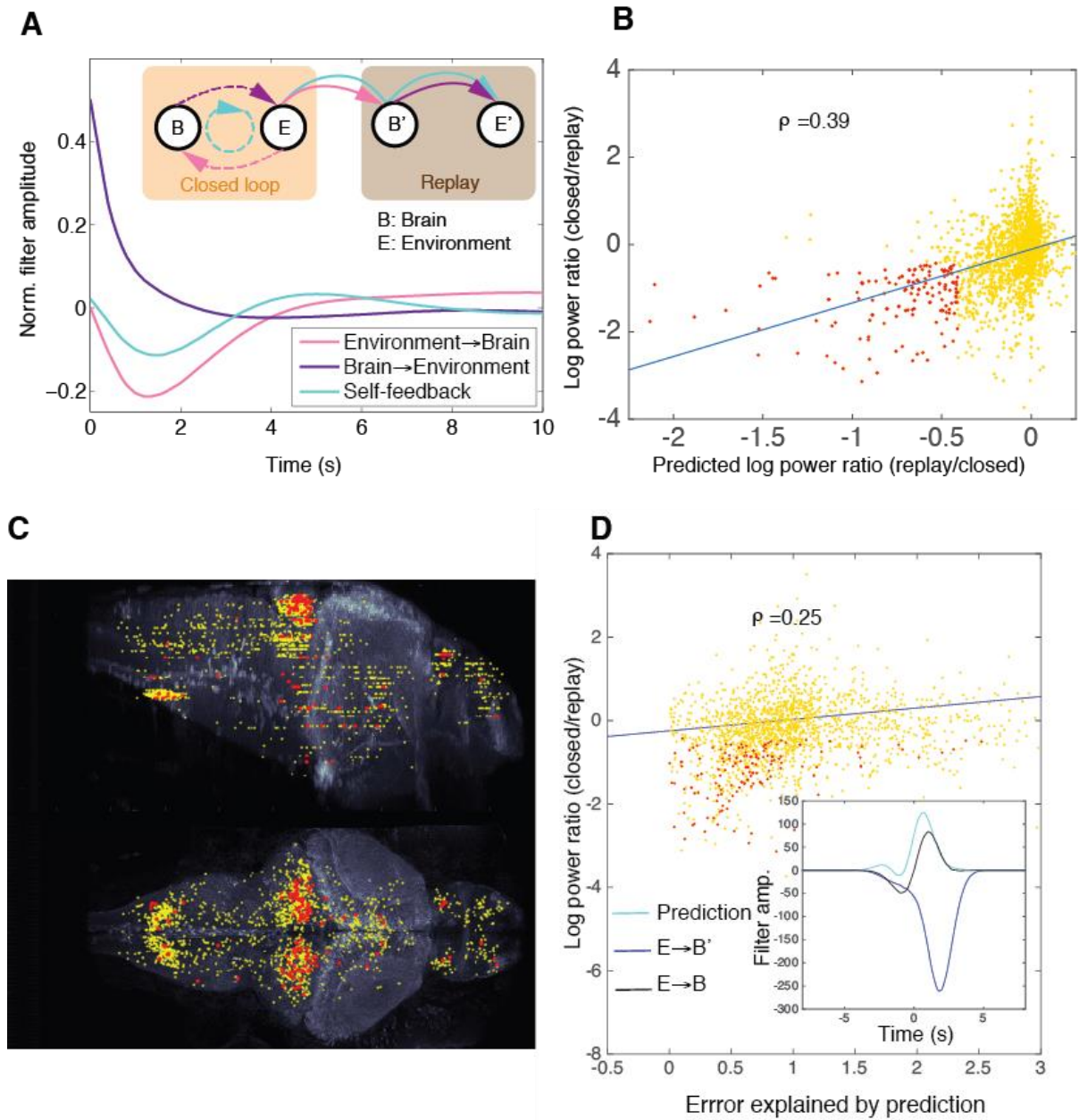

**Figure 3.** CLE feedback predicts suppression of neural fluctuations and correlations. (A) Dynamic interactions were estimated for each neuron by fitting linear filters, whose population averages, after normalizing to peak amplitudes, are summarized. Schematic interactions of the brain (a single neuron) and the environment (motor neuron activity) are shown under the closed-loop and replay conditions (inset). On average, the brain positively drove the environment (B'→E', purple line); the environment negatively drove the brain

(E→B', pink line), and, by combining these two effects, we found that self-feedback (E→B'→E', cyan line) was negative. (B) These filters were then used to predict changes of neural fluctuations under the two conditions. The predicted changes in each neuron based on the filters exhibited strong correlation with the actual changes. Some neurons (top 10%, red dots) exhibited strong negative CLE feedback and were stabilized under the closed-loop condition as predicted by our theory. (C) The location of these neurons are overlaid with the morphology of a reference zebrafish brain (colors as in B). Top panel, side view; bottom panel, top view. Neurons that have strong negative CLE feedback and are strongly stabilized were predominantly located in the cerebellum. (D) The dynamic relation of neuronal activity and motor activity for each neuron in the closed-loop condition (quantified by the E→B filter, naively computed) was qualitatively different from that in the replay condition (the E→B' filter) (inset). For the closed-loop-stabilized cells in (C, red dots), this difference could be explained well by the B'→E' filter from the replay condition (Inset: E→B, black line; E→B', blue line; E→B estimated taking into account of the B'→E' filter from the replay condition, cyan line). The degree to which the difference was accounted for by the B'→E' filter was computed for each cell as a fraction of the mean square error explained. This quantity was positively correlated with the log-power ratio, indicating that the difference in dynamic relation of neuronal activity and motor activity between the closed-loop and replay conditions was greater in the closed-loop stabilized cells.

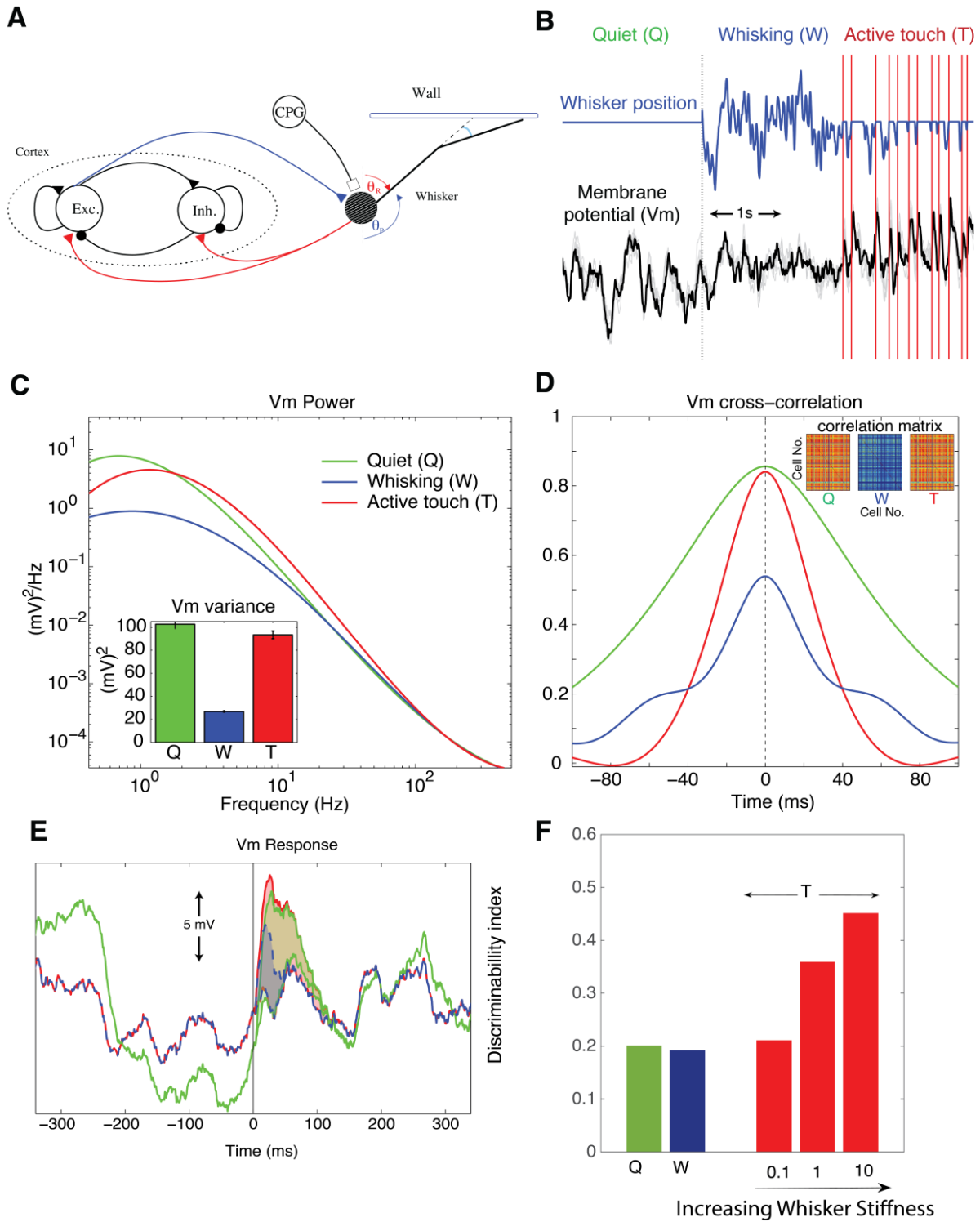

**Figure 4** | A model of the rodent brain state transition. (A) A schematic of the model: 100 excitatory (Exc) and 100 inhibitory (Inh.) neurons receive CLE feedback via a single whisker driven by a central pattern generator (CPG). Triangle and circles represent excitatory and inhibitory synapse respectively. Onset of whisking occurs when the CPG is switched on. CLE feedback is negative overall because the neurons that elicit whisker protraction are assumed

to drive whisker retraction. (B) Membrane potential of cortical neurons (gray lines for individual neurons and black line for population average) and whisker position (blue line) during quiet attentive (Q) whisking (W), and periods of active touch (T). Large and synchronous fluctuations of membrane potential were suppressed during whisking. Active touch elicited reliable responses in these neurons. The vertical dotted line marks the onset of whisking and the vertical red lines mark onset of individual touch event. The power spectrum (C; inset for variance) and cross-correlation (D; inset for correlation matrix of randomly sampled neurons—color warmth indicates the degree of correlation) of membrane potential are averaged over cortical neurons and shown for each condition. Low frequency fluctuations and inter-neural correlation are suppressed during whisking but are recovered during the period of active touch. (E) Membrane potential traces for the Q, W, and T conditions. Sensory events begin at time 0. (F) The discriminability of each type of sensory event. Discrimination performance was similar under Q and W because both signal (exafferent evoked response) and noise (spontaneous fluctuations) were large under Q and both are small during W. Discrimination performance for active touch events were improved relative to the Q and W conditions unless the whisker was too flexible. Discrimination performance was improved with increasing whisker stiffness, reflecting the degree to which CLE feedback was stopped during touch events.

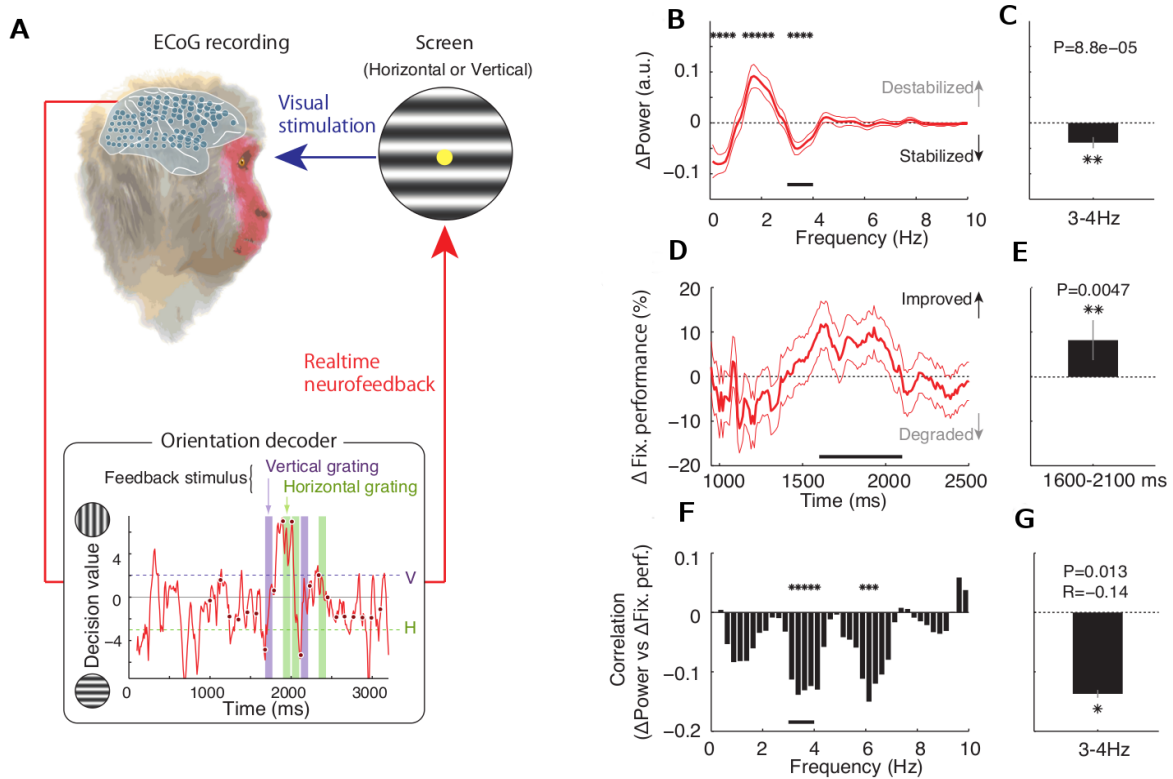

**Figure 5 |** Real-time visual neurofeedback altered brain activity and improved fixation performance. (A) The experimental setup. An estimate of the current visual stimulus (horizontal or vertical grating) is decoded (decision value) every 100 ms from electrodes distributed across the visual cortex. Whenever decision values indicate a high confidence for either grating stimuli the opposite grating stimulus is presented. The stimulus is a gray screen for small decision values (low confidence). The box at the bottom shows the representative decision value dynamics (red trace) and sampling points (circles) used during neurofeedback. The shaded areas show the periods during which the visual stimulus were

presented; the dotted lines, show the decision thresholds (purple: horizontal; green: vertical). Time 0 indicates appearance of the fixation point. (B) Difference of power spectrum of the decoder decision value (Replay-Closed). Average and SEM of whole trials. (C) Summary of the 3-4 Hz amplitude of the difference of decision value in (B). (D) The improvement of fixation performance was quantified by the relative difference between the eyes deviation from fixation point (Replay-Closed) / (Replay+Closed)×2. The mean and SEM of whole trials were used for the analysis. Fixation performance was quantified by the root mean square of the distance from fixation point. (E) Average fixation performance during the time interval [1600-2100 ms] in (D). (F) Trial-to-trial correlation coefficient between the differential fixation performance and the differential amplitude for individual frequencies (bin width: 0.25 Hz). (G) Summary of correlation coefficient for 3-4Hz. The errorbar shows a bootstrapped SEM. Asterisk indicates statistical significance (*: $P < 0.05$; **: $P < 0.005$).

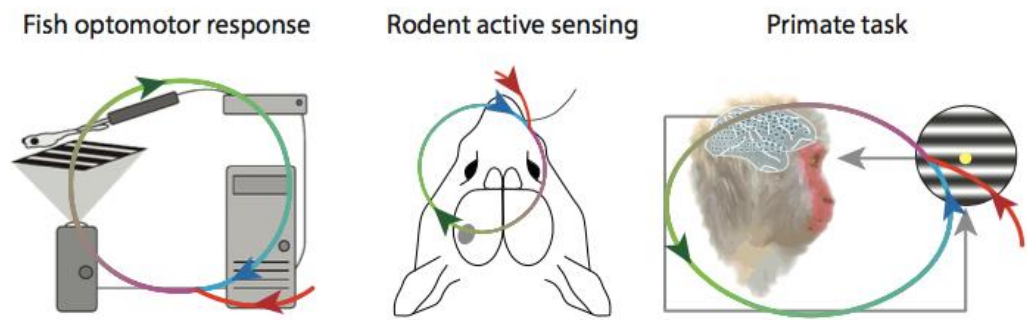

**Figure 6 |** A summary of the three experimental systems studied (Top). In all systems, CLE feedback plays a critical role in determining brain dynamics and behavior, as summarized in the table (Bottom).

**SUPPLEMENTARY INFORMATION**

Contents

**S1 Supplementary fish data**

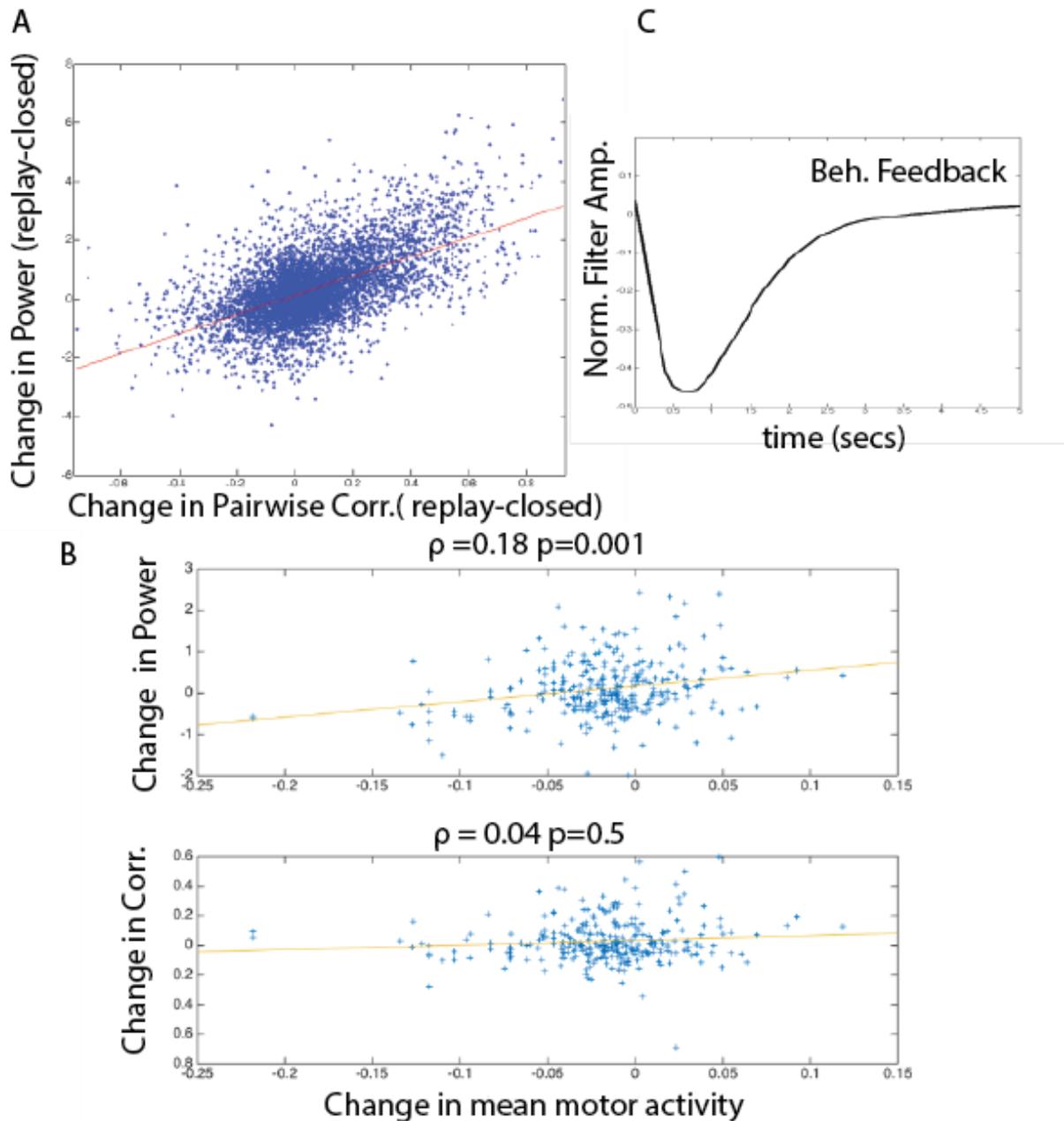

**Figure S1 |** (**A**) The decorrelation effect is not an artifact of measurement noise. Changes in pairwise correlations and change in low frequency power were highly correlated in the recorded neurons even when the each calcium traces (both cells and motor neurons) were thresholded (the threshold was equal to the mean plus one standard deviation of the calcium signal measured over both replay and closed loop conditions) (Spearman's rank correlation $\rho=0.57$, $p<10^{-8}$). (**B**) The increase of motor activity in the closed-loop condition does not explain reduction in neural fluctuations and correlation. (**B**,top): Changes in mean motor activity(replay-closed) and change in low frequency power (mean over interval [0.01 0.15] Hz) per trial (low frequency power is averaged over all cells within a given trial) are positively

correlated (r=0.18, p<10$^{-2}$, Spearman's rank correlation). **B**(bottom): Changes in mean motor activity and changes in pairwise correlations (averaged over all pairwise interactions in a given trial) are not significantly correlated (r=0.03, p>0.5, Spearman's rank correlation). (**C**) Filter describing behavioral feedback**.** A linear filter that describes *behavioral feedback* (E→E') is also strongly negative. Following the kernel method outlined in the methods section we calculated the behavioral feedback as a direct filter between the closed loop environment and environment in the replay condition (E→E'). This also indicates that the CLE feedback to the is strongly negative. Note: the magnitude of the CLE feedback is much greater than the cell self-feedback reflecting the fact that cellular variability was much greater than variability across animals.

## S2 A role for CLE feedback in the brain state transition

In this section, we analyze if sensory input through infraorbital nerve (ION) plays a role in coordinating whisking behavior, thalamic spiking activity, and cortical local field potential (LFP). Previous results based on simultaneous recording from whisker, thalamus, and cortex exhibited that thalamic spiking rate increased and low frequency power of cortical LFP decreased during whisking behavior (Poulet et. al 2012). Here, we reanalyze this data set and quantify temporal coordination between (1) 5-20Hz power of the whisker position, denoted by *Whisker*; (2) thalamic spiking rate computed with 20ms averaging window, denoted by *Thalamus*; (3) and 1-20Hz cortical LFP power, denoted by *Cortex*, recorded from ION-intact animals (n=22) and ION-cut animals (n=19). Raw recordings and these processed traces are shown in Supplementary Figure 2A for an example animal. We chose the 1-20 Hz range for the analysis of cortical LFP power because notable brain-state-dependent changes were previously observed in this range (\cite{poulet14}). The spectrogram was computed using 2s window to reliably estimate the predominant 1Hz power in cortical LFP and the window was gradually shifted in 20ms steps.

Next, we computed cross-correlation functions between these 3 quantities: *Whisker-Thalamus*, *Whisker-Cortex*, and *Thalamus-Cortex*. While the resulting cross-correlation functions were noisy in each animal, a mean cross-correlation function averaged over each animal group exhibited clear common properties. In both ION-intact and ION-cut animals, whisking behavior lead correlated increase in the thalamic activity and decrease in the cortical slow oscillations. Consistent with this result, the thalamic activity was negatively correlated with the low-frequency cortical LFP fluctuations (Supplementary Figure 2B).

However, the position of the mean cross-correlation peaks was significantly shifted in ION-cut animals relative to the ION-intact animals (Supplementary Figure 2B). Specifically, the peak of the *Whisker-Thalamus* cross-correlation was delayed for 400 ms (p=0.02, bootstrap test) and the peak of *Whisker-Cortex* cross-correlation was delayed for 200 ms (p=0.03, bootstrap test) in ION-cut animals. On the other hand, the temporal relationship between the thalamic spiking activity and the low-frequency cortical LFP power was not significantly altered as assessed by the *Thalamus-Cortex* correlation function (p>0.05, bootstrap test). The corresponding bootstrap statistics were computed by randomly resampling animals from the two groups, assuming a null hypothesis that the two animal groups are the same (see, the inset panels for the bootstrap statistics about the difference of the cross-correlation peaks).

These analyses suggest that, after a whisking onset, the brain state transition was delayed in ION-cut animals relative to ION-intact animals. Thus, while sensory input is not necessary for the brain state transition, it was necessary for inducing short-latency brain state transitions.

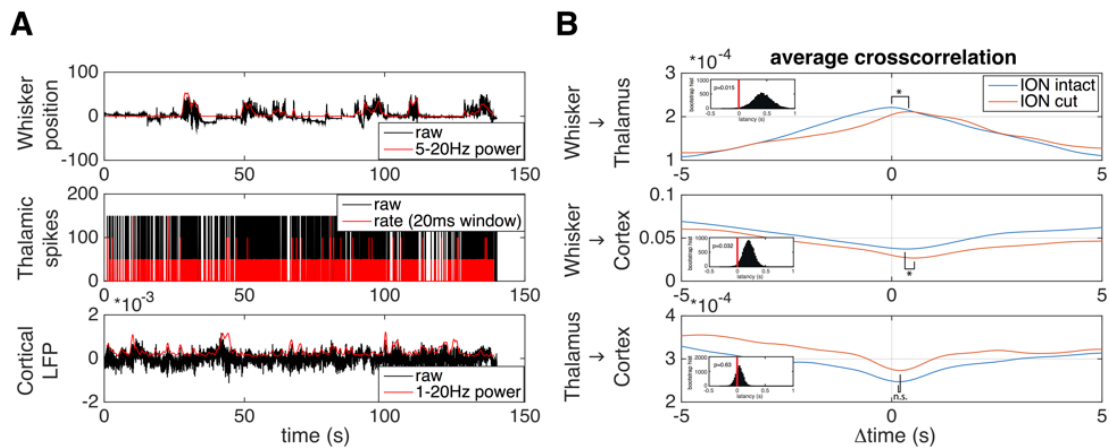

**Figure S2 |** (**A**) A simultaneous recording of whisker position (Top), thalamic spikes (Middle), and cortical LFP (Bottom) in an example animal (Poulet et. al 2012). .Based on these raw traces (black), brain-state-relevant quantities (red) are computed and shown in each panel: 5-20 Hz power of the whisker position (Top), thalamic spiking rate (Middle), 1-20 Hz power of the cortical LFP (Bottom). (**B**) A cross-correlation function between *Whisker* and *Thalamus* (Top), *Whisker* and *Cortex* (Middle), and *Thalamus* and *Cortex* (Bottom) for ION-intact animals (blue) and ION-cut animals (red), where the inset panels show the bootstrap statistics about the difference of the cross-correlation peaks. The *Whisker-Thalamus* and the *Whisker-Cortex* correlation functions were significantly shifted by the ION cut.

S3 Whisker model

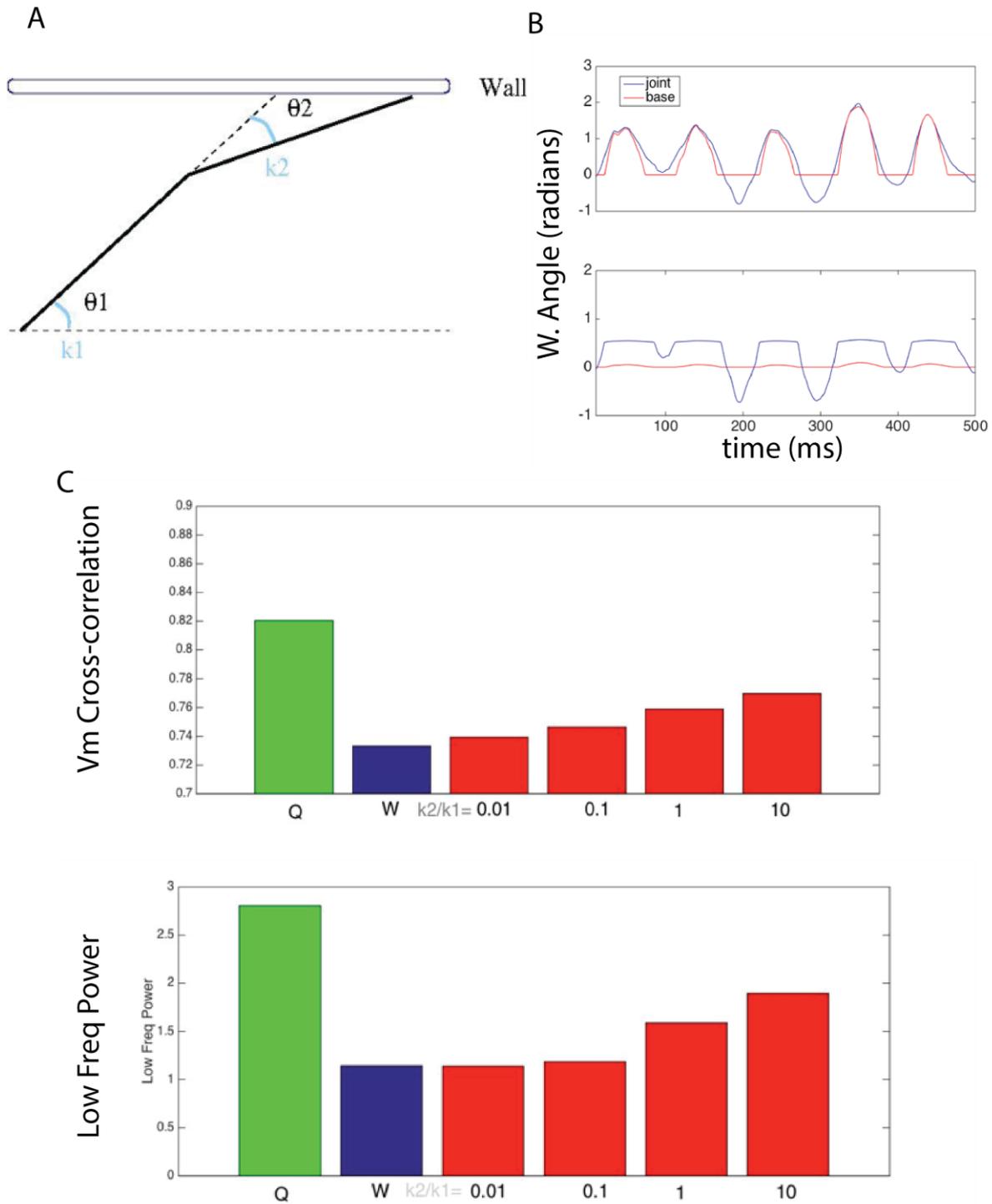

Figure S3: (**A**) A whisker comprising of two sections with a joint angle $\theta_2$ and base angle $\theta_1$. The base and joint are constrained by two springs with spring constants k1, and k2 respectively. Whisking is implemented by driving the equilibrium position of the base spring. The center spring is in equilibrium at zero angular displacement and tends to align the whisker sections. The whisker is length L and massless but constrained by a solid wall placed at y, where y<L. (**B**) The relationship between the joint (red) and base (blue) angle in a flexible, k2 =.1, (middle) and stiff, k2 =10, (bottom) whisker. (**C**) The dependence of cortical fluctuations and responses on whisker stiffness. Model data from quiet attentive (Q, green), whisking (W,blue) and during contact events (red) for whiskers of increasing stiffness. (**C**,top): average cross correlation of membrane between neurons during Q, W and period of active touch for different stiffnesses (k2/(k1=1)). (**C**,bottom): same as top but for low frequency power (average over the range [0.5, 2]Hz).

## S4 A simple conceptual model of CLE feedback

We illustrate our whisker theory with an extension of simple conceptual model presented in the text, Figure 1. It describes the dynamic interaction between brain activity (here membrane potential of neurons) and the body (here whisker position) (Supplementary Figure 4). Specifically a simplified brain-environment system was modeled as a stochastic 1-dimensional first order linear ordinary differential equation as follows,

$$\dot{B}_c(t) = (\omega_s - 1)B_c(t) - \omega E(t) + \xi(t) + I(t)$$

$$E(t) = B_c(t)$$

integrated with a Euler step $dt = 0.01$, where $B_c$ is a dimensionless dynamical variable representing the collective activity of neurons (e.g. local field potential) and $E$ is the environmental variable (e.g. whisker position). Note: for simplicity we assume that the dynamics of the environment variable are much faster compared with the brain dynamics and, thus, E rapidly converges to current B. The negative CLE feedback robustly reduces neural fluctuations unless feedback delay is too large. The parameter, $\omega_s = 0.95$, is the self-coupling within the brain, $\omega$, is the magnitude of the environmental influence on the brain, $I$ is external input and, $\xi$ is unit variance white noise of mean 0. An open-loop condition, i.e., the absence of environmental coupling, is modeled by setting, $\omega = 0$ and the closed-loop condition by setting $\omega = 0.4$. Here we have set $\omega_s$ to be close to criticality ($\omega_s = 1$) to show that even weak CLE feedback can significantly reduce neural fluctuations. However, this choice is not central to our theory because strong CLE feedback works likewise when the system is away from criticality. Passive sensory stimulation was modeled by setting, $I = 1.5$ for a time period of 20. Interruption of the reafferent signal was implemented by fixing the environmental variable for the same period, i.e, setting $E(t') = E(t)$ for $t < t' < t + 20$.

The replay condition is constructed by first recording the environmental variable, $E$, in the closed-loop condition and subsequently replaying this recording to the brain at a different time. The brain activity in the replay condition, $B_r$, is described by

$$\dot{B}_r(t) = (\omega_s - 1)B_r(t) - \omega E(t) + \zeta(t)$$

where $\omega = 0.4$ as in the closed-loop condition and $\zeta$ is white noise during the replay condition.

Like our simple conceptual model under these conditions CLE feedback reduces neuronal gain and suppresses neural fluctuations in the closed-loop condition (Supplementary Figure. 4E, blue lines). Furthermore, the responses of neurons to exafferent input are also suppressed (shaded black interval in Supplementary Figure. 4E). However, large responses during the active state are recovered if exafferent input coincides with brief interruptions of CLE feedback (e.g., in Supplementary Figure. 4E the environmental variable is briefly fixed -- on the surface of an object for whisker touch events -- after the onset of stimulation for a period indicated by the shaded red interval; see Supplementary Figure. 4 for a schematic diagram). Specifically, this brief interruption temporarily increases neuronal gain and enhances the brain's response to exafferent input.

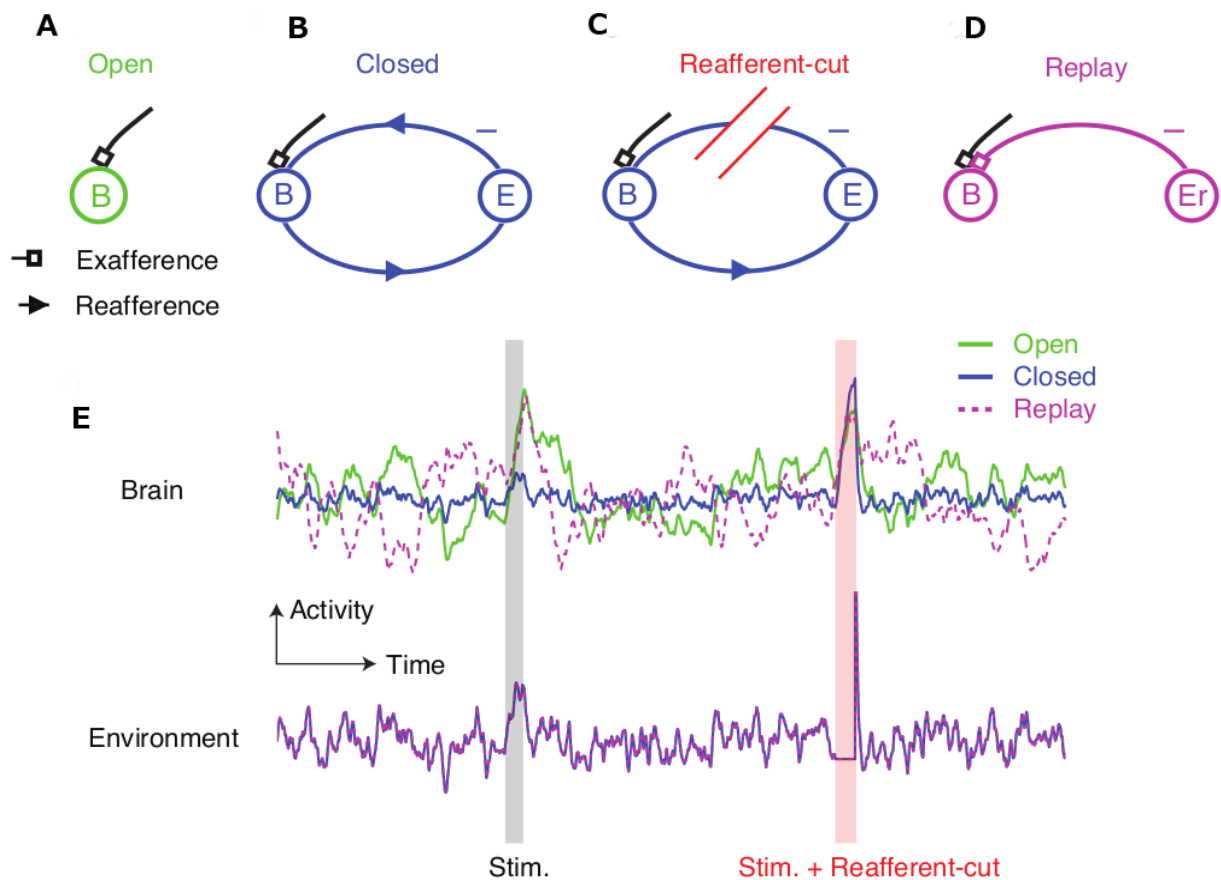

**Figure S4** | CLE feedback explains the changes in brain state at the onset of active behavior. **A-D**: Schematics of the interaction of a brain (B) and environment (E) variable. In the open-loop condition (**A**), the brain receives passive exafferent input only. In the closed-loop condition (**B**), an environment variable mediates negative CLE feedback to the brain. In the reafferent-cut condition, (**C**), this feedback is briefly interrupted. In the replay condition (**D**), the environmental input to the brain in the closed-loop condition is recorded (denoted by Er) and played back to an identical brain (albeit with different noise) at a later time. In all conditions, in addition to the reafferent input, the brain receives a short pulse of exafferent input. (**E**): Traces of the brain and environment variables under different conditions. In the open loop condition (green) the brain shows large spontaneous fluctuations and large response to exafferent stimulation, which are suppressed during the closed-loop condition (red). The gray bar denotes periods of stimulation. In contrast, the brain variable exhibits large and reliable response to the same input when combined with a brief interruption of the

environment (the pink bar denotes the period of stimulation and reafference interruption). In the replay condition (magenta), the brain behaves similarly to the open-loop condition despite receiving the same input from the environment as the closed-loop condition.

**S5 Decoding of visual stimulus based on ECoG signals**

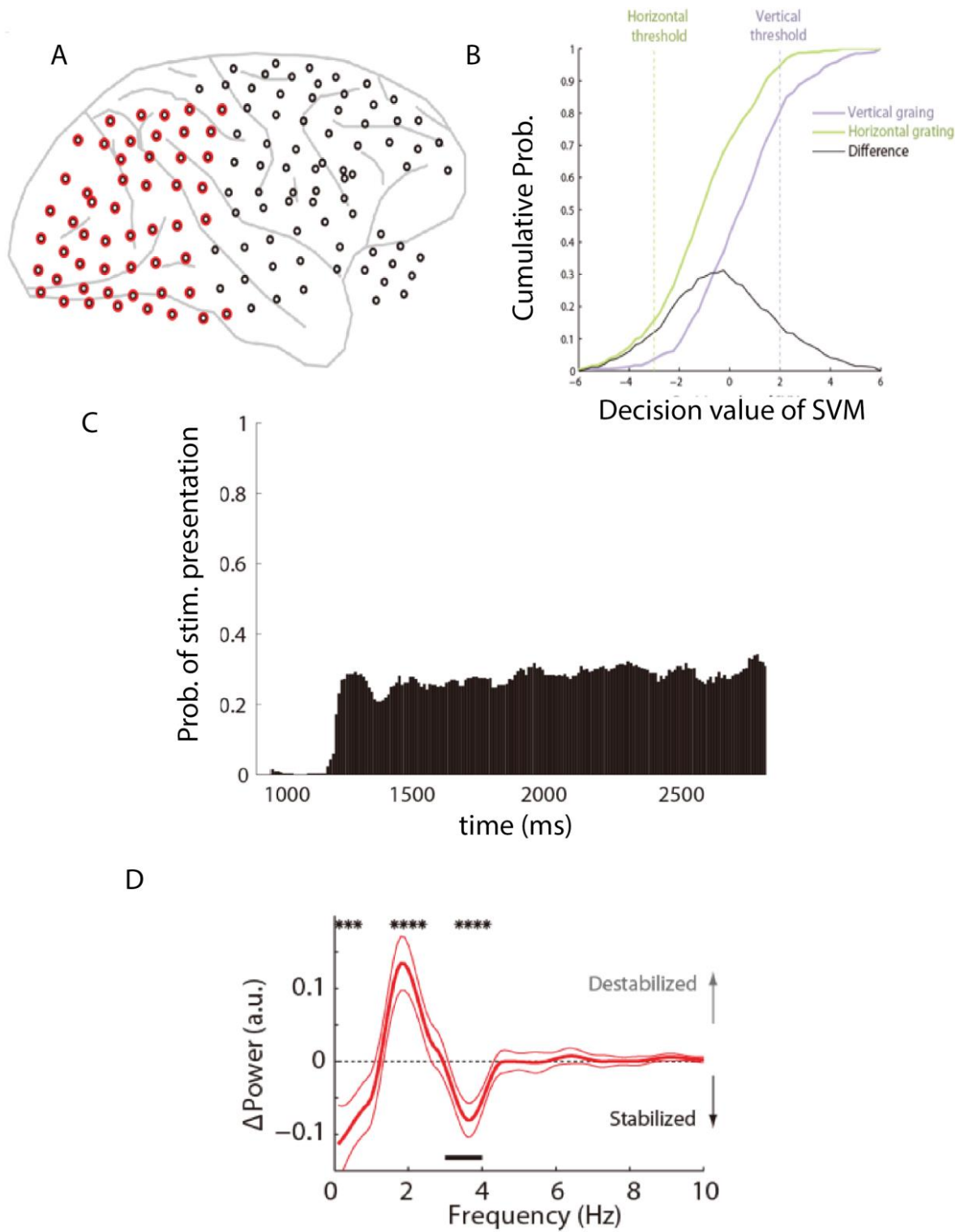

**Figure S5 |** Decoding of visual stimulus based on ECoG signal. (**A**) ECoG electrodes used in the visual neurofeedback experiment. Small black circles indicate the loci of electrodes. The electrodes surrounded by red circles were used for decoding by SVM. (**B**) Decision value distribution for a trained SVM, in the visual neurofeedback experiment. The purple and the green curves indicate cumulative probability distribution of the decision value during presentation of vertical and horizontal gratings, respectively. The dashed lines show the criteria for presenting feedback stimulus. (**C**) Averaged temporal evolution of stimulus presentation frequency within a trial during the visual neurofeedback experiment. The visual stimulus started to appear typically 1300 ms after the onset of fixation point. To avoid presenting stimulus before monkey fixated its eyes, no stimulus was presented 0-1000 ms after the appearance of fixation point. (**D**) Difference of power spectrum of the decoder decision value (Replay-Closed) during early period (before the end of behavioral improvement, 1000-2100 ms). The trend was not changed from the result for whole stimulus-presentation period (1000-2800 ms), including the significant reduction of the 3-4 Hz power by the neurofeedback. Convention follows Figure 4B. The black bar indicates the 3-4 Hz, used in the analysis of behavioral correlation in Figure 4f and 4G.

**S6 Alternate schemes for sensory feedback in the whisker system**

A

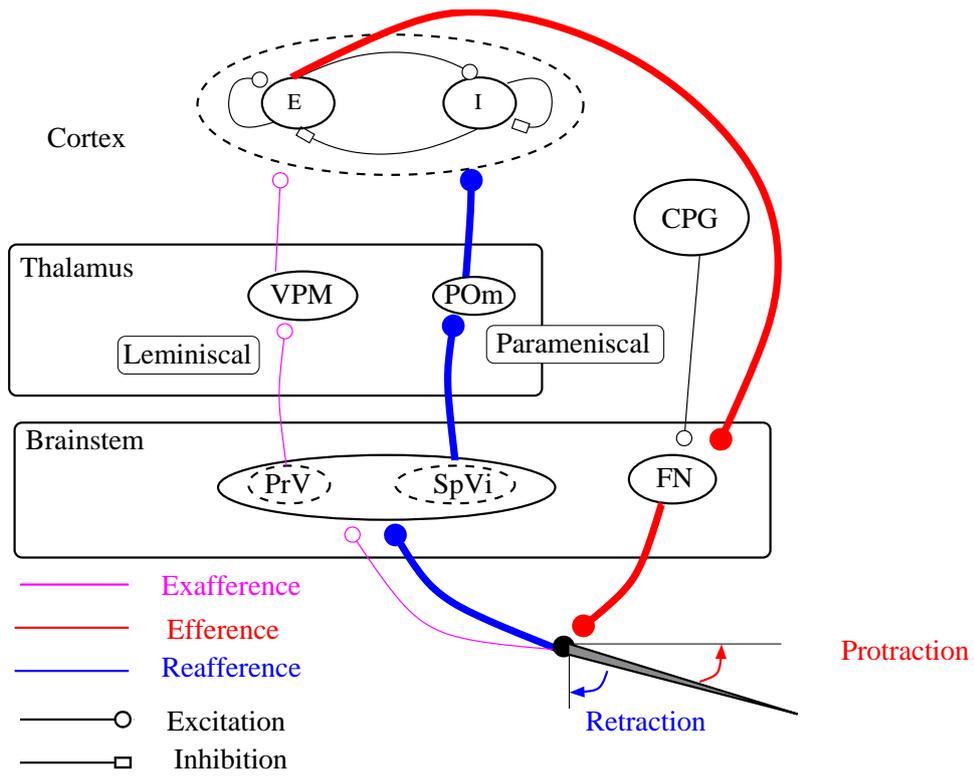

B

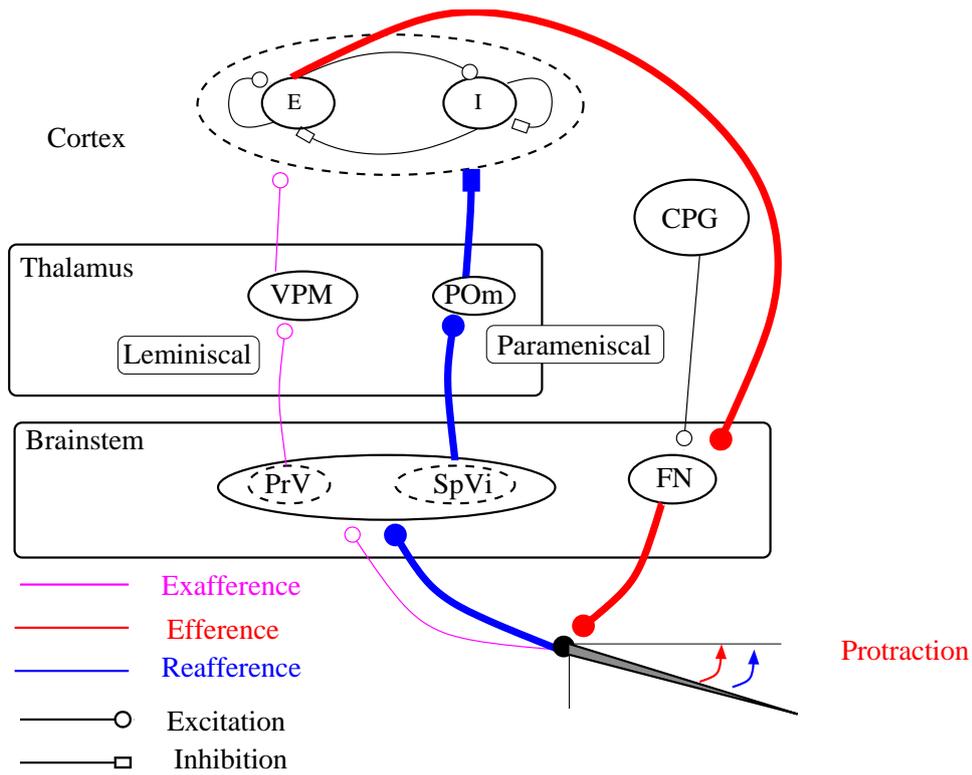

C

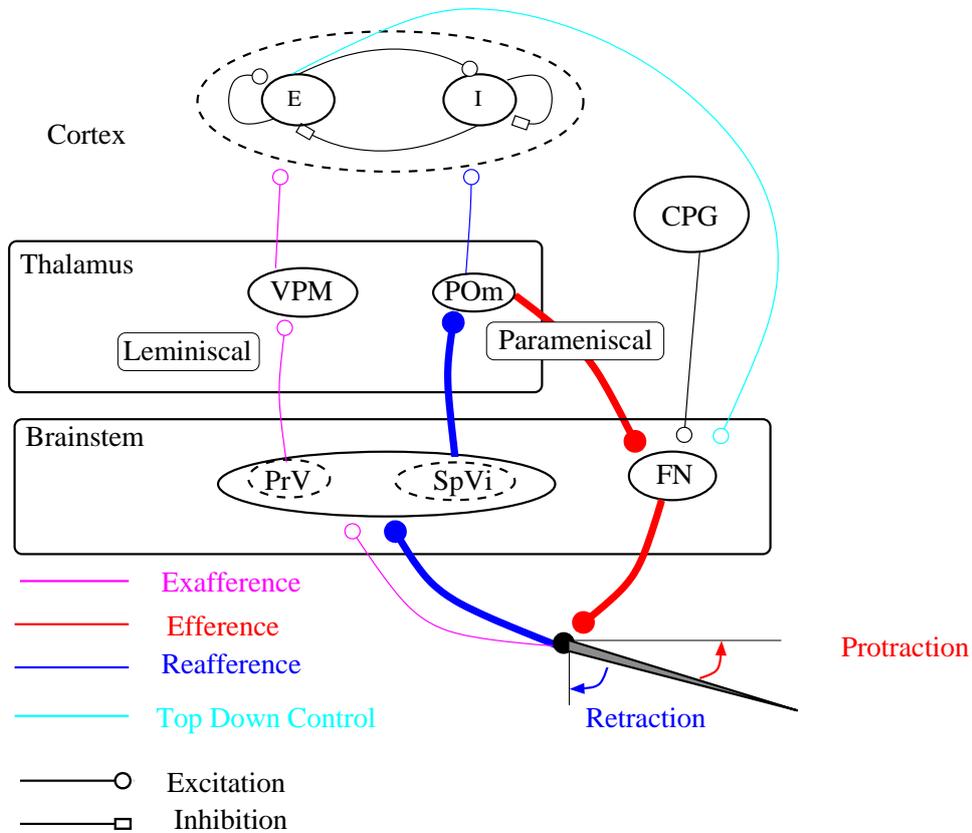

**Figure S6**: Three models for whisker circuits mediating a negative CLE feedback. (**A**) Net activation of the modeled cortical population drives neurons in the facial nucleus (FN) to drive

whisker protraction. This in turn reduces excitatory sensory input to the modeled cortical population because this is driven by retraction. Here negative CLE feedback is mediated implicitly at the periphery. (**B**) Protraction information could be conveyed along the full pathway but net inhibitory input to the modeled cortical population result because POm inhibits the cortex. In either model circuit, the initial activation of cortical neurons causes subsequent suppression of their activity by feedback through the whisker circuit, constituting a negative sensory feedback loop. These two hypotheses are testable but not necessarily mutually exclusive. (**C**) An alternative hypothesis is the whisker feedback is completed in the brain stem and changes in cortical activity are driven by activity changes in the thalamus.

Our whisker model remains abstract in terms of known vibrissa system anatomy and, in particular, the relay stations between the cortex and a whisker. The exact concordance of the model with known vibrissa system anatomy is beyond the scope of this paper, but we provide a more detailed to demonstrate a possible anatomical explanation of our model and provide a means for the research community to experimentally examine CLE feedback in specific biological circuits.

In our model (Supplementary Figure 6A), we assume that projections between regions are largely excitatory (c.f. Ahissar 2010). Importantly, we distinguish two subcortical pathways that signal afferent input to cortical neurons - one for transmitting reafferent input and the other for transmitting exafferent input. This distinction could reflect the separation between a parameniscal pathway i.e., via thalamic POm, conveying reafferent signals, and a lemniscal pathway, i.e., via thalamic VPM, conveying exafferent input (Pierret et al., 2000; Urbain et al., 2015). Accordingly, we modeled exafferent input to cortical neurons by using a stereotypical pulse upon each whisker contact and brief deflection, and reafferent input proportional to whisker angle reflecting motor efference (Szwed et al., 2003). Regardless of how the properties of reaffererent input - whisking phase, absolute position, or their temporal derivatives - are encoded by the pathway they do not change the main conclusion of our model as long as the CLE feedback constitutes net negative feedback.

In agreement with the anatomy, we assume that cortically generated motor signals modulate whisking behavior by acting on the facial nucleus (FN) (Ahissar, 2010). Because whisking behavior persists after sensory denervation (Welker, 1964), cortical ablation (Semba and Komisaruk, 1984), or decerebration (Lovick, 1972), we explicitly modeled a central pattern generator (CPG) that autonomously generates whisking patterns locating exogenous to the cortical-whisker loop (Hill et al., 2011). Thus, the FN receives input from both the cortical population and CPG and moves the whisker in the reafference model.

We cannot rule out other biological pathways for negative CLE feedback. For example, negative feedback can also be mediated by dominant cortical inhibition to the modeled population of cortical neurons (Supplementary Figure 6B). In agreement, thalamocortical connections strongly innervate fast spiking neurons and consequently implement strong feedforward inhibition to the cortex (Bruno and Simons 2002). Other potential models arise from heterogeneity in cortical populations. For example, negative CLE feedback can be mediated by neurons in the barrel cortex that directly drive whisker retraction with extremely short latencies (Matyas et al., 2010).

Alternatively, the dominant negative feedback loop could be subcortcial, Supplementary Figure 6C. Here the dynamics of the cortex only indirectly reflects the stabilization of the thalamus. This scheme is also consistent with reduced thalamic activity during whisking (Poulet et al., 2012). It is important to note that this implementation also fundamentally relies on stabilisation of neuronal activity by negative CLE feedback.

At the level of the whole vibrissa system, there are likely multiple parallel and nested feedback loops, both positive and negative (Ahissar et al., 2003). However, we assume that the overall or net feedback mediated by the cortical-whisker circuit during corresponding behavior is negative in sign which we empirically demonstrate the presence of negative CLE feedback in zebrafish active sensing. These models are experimentally testable. For example, a group of neurons that encode aspects of reafferent input, such as whisker protraction, could be genetically labeled and used for anatomical tracing studies. We can moreover study the physiological role of these neurons by optogenetically silencing them during active whisking and study how brain state as well as the animal's behavior may be altered. This specific neuronal population could also be optogenetically activated and the ensuing behaviors and changes in vibrissae information processing pathways studied.